# *Heteronuclear transfers from labile protons in biomolecular NMR: Cross Polarization, revisited*

Mihajlo Novakovic[1], Sundaresan Jayanthi[2], Adonis Lupulescu[3], Maria Grazia Concilio[1], Jihyun Kim[1], David Columbus[1], Ilya Kuprov[4], and Lucio Frydman[1*]

[1]Department of Chemical and Biological Physics, Weizmann Institute of Science, Rehovot 7610001, Israel

[2]Department of Physics, Indian Institute of Space Science and Technology, Valiamala, Thiruvananthapuram 695 547, Kerala, India

[3]Nicolae Titulescu nr. 8, Turda, Jud. Cluj, Romania

[4]School of Chemistry, University of Southampton, Southampton SO17 1BJ, UK

*Email: lucio.frydman@weizmann.ac.il

**Abstract**

INEPT- and HMQC-based pulse sequences are widely used to transfer polarization between heteronuclei, particularly in biomolecular spectroscopy: they are easy to setup and involve low power deposition. Still, these short-pulse polarization transfers schemes are challenged by fast solvent chemical exchange. An alternative to improve these heteronuclear transfers is J-driven cross polarization (J-CP), which transfers polarization by spin-locking the coupled spins under Hartmann-Hahn conditions. J-CP provides certain immunity against chemical exchange and other $T_2$-like relaxation effects, a behavior that is here examined in depth by both Liouville-space numerical and analytical derivations describing the transfer efficiency. While superior to INEPT-based transfers, fast exchange may also slow down these J-CP transfers, hurting their efficiency. This study therefore explores the potential of repeated projective operations to improve $^1H \rightarrow {}^{15}N$ and $^1H \rightarrow {}^{15}N \rightarrow {}^{13}C$ J-CP transfers in the presence of fast solvent chemical exchanges. It is found that while repeating J-CP provides little $^1H \rightarrow {}^{15}N$ transfer advantages over a prolonged CP, multiple contacts that keep both the water and the labile protons effectively spin-locked can improve $^1H \rightarrow {}^{15}N \rightarrow {}^{13}C$ transfers in the presence of chemical exchange. The ensuing Looped, Concatenated Cross Polarization (L-CCP) compensates for single J-CP losses by relying on the $^{13}C$'s longer lifetimes, leading to a kind of "algorithmic cooling" that can provide high polarization for the $^{15}N$ as well as carbonyl and alpha $^{13}Cs$. This can facilitate certain experiments, as demonstrated with triple resonance experiments on intrinsically disordered proteins involving labile, chemically exchanging protons.

## Introduction

Modern NMR relies on experiments that correlate different nuclei, and spread their chemical shift information into multiple dimensions.[1,2] These correlations are usually performed using coherent polarization transfer schemes; for many protein and nucleic acid experiments these schemes start from $^1$H, and utilize J-couplings to transfer polarizations or coherences to bound $^{15}$N –and, occasionally, onwards to $^{15}$N-bound $^{13}$C. Besides improving resolution and enriching the spectral information, these heteronuclear transfers are important for enhancing sensitivity:[3–6] Maximizing the signal-to-noise ratio, particularly in the context of indirect low-γ nuclei detection, dictates that most experiments start with the excitation –and if feasible end with the detection– of $^1$H. Improving the transfer of spin order among nuclei with different γ, is thus of great importance in NMR. Two different approaches are normally used to achieve such transfers. The "pulse-interrupted free precession" category[4] includes relatively long periods of RF-free evolution, and encompasses the coherence transfer blocks used in INEPT-,[5,7] DEPT-,[8] HSQC-,[9,10] and HMQC-based[6,11,12] experiments. Such sequences are easy to set up, robust, and have low power depositions. Still, by virtue of their reliance on multiple lengthy free precession periods, these transfers can significantly degrade in the presence of fast chemical exchange between the originating $^1$H and an aqueous solvent.[13–15] A second approach to heteronuclear transfers is J-driven cross-polarization (J-CP)[16,17] or, more generally, "spin-order transfer under an average Hamiltonian".[18] J-CP actually predates INEPT;[5,16] in general it involves achieving, via the application of appropriately designed irradiations, an effective Hamiltonian possessing double- or zero-quantum terms that do not commute with the starting I ($^1$H) or S ($^{15}$N) spin operators. Letting the initial spin state evolve under the influence of such average Hamiltonian can then transfer spin order among the *J*-coupled spins.[19,20] The simplest example of this arises when transferring polarization by simultaneously applying continuous spin-locking fields $B_1$ on an $^1$H-$^{15}$N pair, satisfying the $\omega_1 = \gamma_H B_{1H} = \gamma_N B_{1N}$ Hartmann-Hahn condition[21]. This leads to an $H_x \rightarrow N_x$ transfer that is mediated by a zero-quantum state;[20] in an ideal, relaxation-free transfer, this would be described by

$$H_X \rightarrow H_X \frac{(1+\cos \pi J t)}{2} + N_X \frac{(1-\cos \pi J t)}{2} + 2(H_Y N_Z - H_Z N_Y)\frac{\sin \pi J t}{2}, \quad (1)$$

where *J* represents the $^1$H-$^{15}$N scalar coupling, and a usual spin-1/2 operator notation is used.[22]

Because of its sensitivity to the matching settings and to spatial RF inhomogeneities, plus its imposition of higher RF heating loads (particularly on physiological solutions at high-fields), CP –which is widely used in solid state NMR[23,24]– is less common in liquids than the INEPT-related approaches. J-CP's spin-locking, however, imparts an immunity against chemical exchange and relaxation effects, that makes it a potential method of choice to transfer polarization in disordered proteins,[15,25] or for the fast-exchanging amino/imino moieties of nucleic acids.[14] Furthermore, multiple CP schemes including MLEV,[26,27] DIPSI,[28] WALTZ[29] have been developed to alleviate the aforementioned inhomogeneity and heating-derived problems.[17,18,30–34] Still, even these schemes become inefficient when the exchange rate $k_{HW}$ of the $H^N$ with water, exceeds $\approx 10J$. The reasons for this are not immediately evident, considering that abundant solvent water protons $H^{water}$ are constantly repolarizing the $H^N$ bound to the $^{15}N$. Furthermore, the relatively long spin relaxation times of $^{15}N$ (and $^{13}C$) in such systems,[35,36] could enable principles akin to those used in the recently proposed Looped, projected spectroscopy (L-PROSY)[37] experiment, to improve these heteronuclear transfers even in fast chemical exchange cases. In view of these issues the present study revisits J-CP transfers in exchanging systems, considers a number of J-CP alternatives derived from L-PROSY,[37,38] and shows how, under certain conditions, these looped approaches can be used to increase the efficiency of heteronuclear polarization transfers in solvent-exchanging biomolecules.

## Results

### The $H^N \xrightarrow{J_{NH}} N^H$ polarization transfer step in the presence of chemical exchange

We focus on a system involving a labile $H^N$ bonded to an $^{15}N$, which exchanges with the aqueous protons $H^{water}$. To compare the extents to which the exchange rate $k_{HW}$ will affect J-CP vs INEPT, Liouville-space simulations based on Spinach codes were run.[39–41] These took into account either a perfectly matched Hartmann-Hahn process, or an idealized refocused INEPT process. The simulations solved the Liouville - von Neumann equation

$$\frac{\partial \rho}{\partial t} = -i[\mathcal{H}, \rho(t)] + \hat{R}\rho(t) + \hat{\Xi}\rho(t), \qquad (2)$$

where the double rotating-frame Hamiltonian

$$\mathcal{H} = \Omega_{HN}H_Z + \Omega_{HW}W_Z + \Omega_N N_Z + 2\pi J H_Z N_Z + \omega_{1H}(H_X + W_X) + \omega_{1N}N_X \qquad (3)$$

contained Zeeman terms for the exchangeable $H^N$ ($H$) and for $H^{water}$ protons ($W$) as well as for the $N^H$ nitrogen ($N$), a scalar J-coupling between $H^N$ and $N^H$, and a continuous $\omega_1 = \omega_{1H} = \omega_{1N}$ irradiation along the x-axes. $\hat{\Xi}$ in Eq. (2) is a superoperator[42,43] describing the chemical exchanges between $H^N$ and $H^{water}$ at rates weighted by these ensembles populations; $\hat{R}$ is a superoperator accounting for relaxation within an extended $T_1/T_2$ approximation, wherein product states relax at the sum of their constituents' relaxation rates. The effects of exchange on the spin-lattice relaxation rate in the rotating frame ($R_{1\rho}$) arising during the CP field were accounted *ad hoc,* by considering that for effective nutation fields governed by the proton offset $\Omega$ (which will differ for $H^{water}$ and for $H^N$) and for a given $\omega_1$ spin-lock field, $R_{1\rho}$ will be given by[44,45]

$$R_{1\rho} = R_1 cos^2\Theta + R_2 sin^2\Theta + sin^2\Theta \frac{p_1 p_2 \Delta\omega^2 k_{HW}}{k_{ex}^2 + \omega_1^2 + \Omega^2} \qquad (4)$$

where $\Theta = \arctan(\omega_1/\Omega)$, $R_1$ and $R_2$ are longitudinal and transverse relaxation rates, $k_{HW}$ is the chemical exchange rate between $H^N$ and water, $p_1$ and $p_2$ are the populations of $H^N$ and $H^{water}$ respectively, and $\Delta\omega$ is the chemical shift difference between the two $^1H$ pools. INEPT experiments were similarly simulated in Spinach, under the ideal-pulses assumption.

With these tools, a number of $H^N \rightarrow N^H$ polarization transfer scenarios were considered. Skrynnikov and coworkers[15] observed that refocused INEPT transfer efficiency drops to ca. 20% if $k_{HW} \sim 2J$ and becomes impractical if exchange rates are $>3J$; they also reported a better performance for J-CP, with $H^N \rightarrow N^H$ transfer efficiencies dropping to 20% only when $k_{HW} > 10J$. All this is reproduced by the Spinach-based simulations (Figure 1a); it is also corroborated by experimental L-alanine observations recorded for its amino group (Figure 1b). J-CP's extended efficiency can be rationalized by the fact that, if spin locked, water protons will constantly replenish the amide sites with $H_X$ polarization. Although this will not preserve the zero-quantum states involved in the CP process (Eq. (1)), it will partly offset the effects of the exchange. Eventually, however, it seems the $^1H \rightarrow ^{15}N$ polarization transfer becomes considerably attenuated by exchange even when using J-CP; this, despite the availability of a nearly endless supply of solvent-based polarization.

To clarify this last point, it is enlightening to consider the changes that $k_{HW}$ imparts on the sinusoidal J-CP oscillations described by Eq. (1). Figure 2a explores this with the aid of Liouville-space simulations, focusing on the buildup that the $^{15}N$ polarization will undergo during the J-CP processes when acted by a chemical exchange. For zero or small $k_{HW}$ rates, the coherent $(1 - \cos \pi Jt)$ J-driven oscillations predicted by Eq. (1) – slightly damped due to the effects of $^1H$ and $^{15}N$ $T_{1\rho}$ relaxation– are dominant. At higher $k_{HW}$ values, the amplitude of these *J*-oscillations is reduced and, as described by Wong et al,[46] the conditions needed for achieving maximal $^{15}N$ polarization move to longer times. Eventually, for $k_{HW} > 5J$, an oscillation-free buildup is observed. Supporting Figure S1 compares this $^{15}N$ behavior against the concurrent proton depletion happening during J-CP, where a similar change can be observed as well. Experimental results on a slightly different $^{15}N$-$^1H_3$ system (Figure 2b) confirm this behavior, with $^{15}N$ CP buildups measured on $^{15}N$-labeled alanine at different temperatures illustrating the attenuation of the J oscillations produced by the chemical exchange.

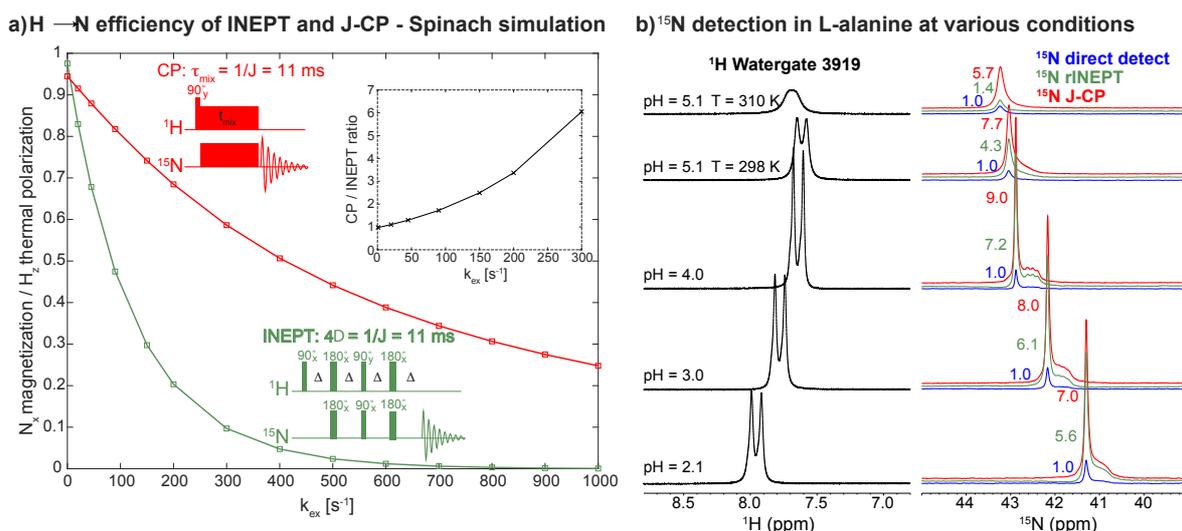

**Figure 1**. a) Simulations comparing the efficiencies of refocused INEPT (green) and of J-CP (red) transfers, as a function of chemical exchange rate between exchangeable proton and solvent water. CP's robustness over INEPT is further stressed in the inset, which shows the efficiency ratio between the two methods increasing rapidly with $k_{HW}$. 100 water protons were used in these simulations, a much larger spin pool than $H^N$. $T_{1N}$ and $T_{2N}$ were estimated at 0.8 s; $^1H$ relaxation parameters were: $T_{1w}$ = 3s, $T_{2w}$ = 0.5s, $T_{1HN}$ = 0.6 and $T_{2HN}$ = 0.2s. b) $^1H$- and $^{15}N$-detected experiments on L-alanine's amino group at different pH conditions and temperatures –both of which change the rate of $^1H$ solvent exchanges. This is manifested by the different J-splittings and linewidths in the $^1H$ resonance. As predicted by simulations, a more sensitive $^{15}N$ detection can be achieved with CP than with INEPT in the presence of chemical exchange. The broad hump present in $^{15}N$ spectra –noticeable as a multiplet structure at pH = 4– arises from J-couplings between nitrogen and deuterium, which is present in the sample for lock (10% $D_2O$). Numbers next to the peaks represent peak intensity relative to $^{15}N$ direct detection (acquired with long interscan delay). Spectra were acquired at 1 GHz using a Bruker Avance Neo spectrometer equipped with a TXO cryoprobe.

The exchange-driven changes shown in Figure 2 are reminiscent of changes reported by Ernst *et al*, in early $^1H \rightarrow ^{13}C$ ferrocene single crystal CP experiments.[49] In such solid NMR cases CP was mediated by the $^{13}C$-$^1H$ one-bond dipolar coupling, and it was homonuclear couplings between the latter and remote $^1H$s, that dampened the CP oscillations. Simplified models involving a coherent interaction within the dipole-coupled spins and an incoherent spin-diffusion within the $^1H$ bath were used to reproduce this behavior,[49,50] leading in general to an exponential transfer buildup reaching a full $\gamma_H/\gamma_C$ enhancement. To the best of our knowledge no comparable theory has been developed for the solution-state case in Figures 1-2, and the asymptotic limits reached by J-CP in solvent exchanging systems have not been hitherto reported; we thought it valuable to describe these here. The simplified system of differential equations used to represent J-CP in such exchanging liquids was

$$\frac{d}{dt}\begin{pmatrix}\langle W_X\rangle\\ \langle H_X\rangle\\ \langle N_X\rangle\\ \langle 2H_YN_Z\rangle\\ \langle 2H_ZN_Y\rangle\end{pmatrix}=\begin{pmatrix}-R_W-k_{WH} & k_{HW} & 0 & 0 & 0\\ k_{WH} & -R_H-k_{HW} & 0 & -\pi J/2 & \pi J/2\\ 0 & 0 & -R_N & \pi J/2 & -\pi J/2\\ 0 & \pi J/2 & -\pi J/2 & -R_H-k_{HW} & 0\\ 0 & -\pi J/2 & \pi J/2 & 0 & -R_N-k_{HW}\end{pmatrix}\begin{pmatrix}\langle W_X\rangle\\ \langle H_X\rangle\\ \langle N_X\rangle\\ \langle 2H_YN_Z\rangle\\ \langle 2H_ZN_Y\rangle\end{pmatrix} \quad (5)$$

where $\langle W_X\rangle$, $\langle H_X\rangle$ and $\langle N_X\rangle$ are expectation values of the corresponding in-phase single-quantum operators for the water $^1$Hs, the labile proton and the nitrogen respectively. As J-CP is mediated by zero-quantum states, $2H_YN_Z$ and $2H_ZN_Y$ coherences are also involved in the description; $\langle 2H_YN_Z\rangle$ and $\langle 2H_ZN_Y\rangle$ in Eq. (5), denote the expectation values of these coherences. All in-phase single-quantum coherences are assumed to be subject to an ideal spin-lock along the *x*-axis, involving a perfect Hartmann-Hahn match free from chemical shift effects. The only parameters involved are then the scalar J N-H coupling, the forward and reverse chemical exchange rates $k_{HW}$ and $k_{WH}$ between labile and water protons, and rotating-frame relaxation times $R_W$, $R_H$ and $R_N$ for the three species. Describing the spin dynamics emerging from Eq. (5) requires deriving expressions for the eigenvalues and eigenvectors of its generator matrix. In view of the complexities of such treatment, analytical expressions for the dynamics were derived using a series of approximations. The lack of direct *J* couplings and the large size of its reservoir, allowed us to assume that the evolution of water's magnetization is mainly governed by transverse relaxation. This leads to a simplified system which includes only $H^N - N^H$ evolution plus a water term that acts as source for $H_X$'s repolarization. Equation (5) then becomes

$$\frac{d}{dt}\begin{pmatrix} \langle H_X \rangle \\ \langle N_X \rangle \\ \langle 2H_Y N_Z \rangle \\ \langle 2H_Z N_Y \rangle \end{pmatrix} = \begin{pmatrix} -R_H - k_{HW} & 0 & -\pi J/2 & \pi J/2 \\ 0 & -R_N & \pi J/2 & -\pi J/2 \\ \pi J/2 & -\pi J/2 & -R_H - k_{HW} & 0 \\ -\pi J/2 & \pi J/2 & 0 & -R_N - k_{HW} \end{pmatrix}\begin{pmatrix} \langle H_X \rangle \\ \langle N_X \rangle \\ \langle 2H_Y N_Z \rangle \\ \langle 2H_Z N_Y \rangle \end{pmatrix}$$

$$+ k_{WH}\begin{pmatrix} W_X^0 exp(-R_W t) \\ 0 \\ 0 \\ 0 \end{pmatrix} \tag{6}$$

where $W_X^0$ represents water's thermal polarization. Equation (6) still doesn't lead to analytical solutions conveying a simple physical picture; to obtain these a further approximation was introduced, by replacing the individual $H^N$ and $^{15}N$ relaxation rates by an effective, common rotating-frame relaxation rate $R_{eff} = (R_H + R_N)/2$. Equation (6) then becomes

$$\frac{d}{dt}\begin{pmatrix} \langle H_X \rangle \\ \langle N_X \rangle \\ \langle 2H_Y N_Z \rangle \\ \langle 2H_Z N_Y \rangle \end{pmatrix} = \begin{pmatrix} -R_{eff} - k_{HW} & 0 & -\pi J/2 & \pi J/2 \\ 0 & -R_{eff} & \pi J/2 & -\pi J/2 \\ \pi J/2 & -\pi J/2 & -R_{eff} - k_{HW} & 0 \\ -\pi J/2 & \pi J/2 & 0 & -R_{eff} - k_{HW} \end{pmatrix}\begin{pmatrix} \langle H_X \rangle \\ \langle N_X \rangle \\ \langle 2H_Y N_Z \rangle \\ \langle 2H_Z N_Y \rangle \end{pmatrix}$$

$$+ k_{WH}\begin{pmatrix} W_X^0 exp(-R_W t) \\ 0 \\ 0 \\ 0 \end{pmatrix}. \tag{7}$$

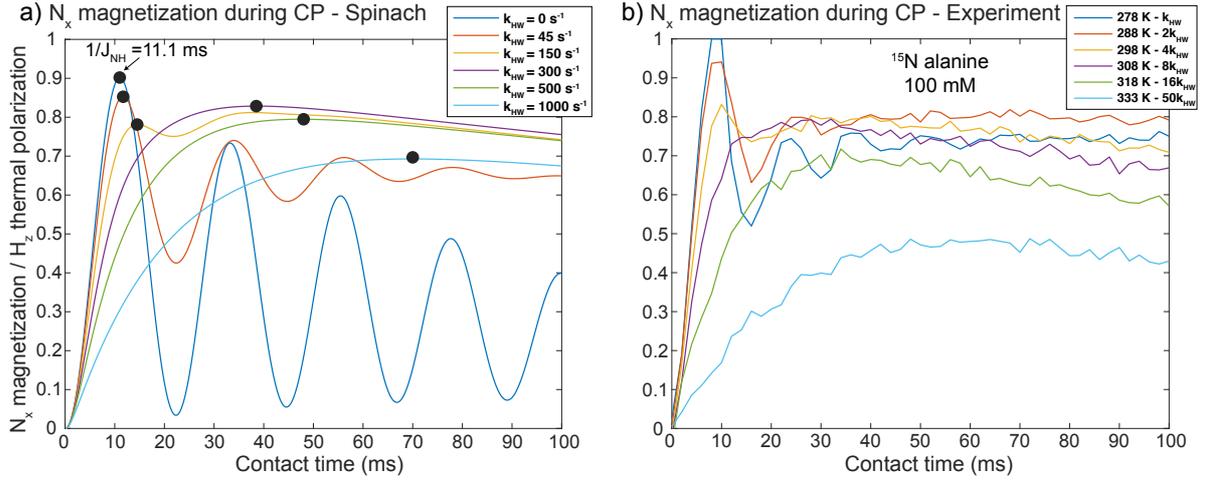

**Figure 2.** a) Buildup of nitrogen $N_X$ magnetization upon cross-polarization for different $k_{HW}$ rates between water and the labile proton. 200 water protons were used in these Spinach simulations –a sufficiently large spin pool vs the solute (see Supporting Information 4). Buildups were computed up to 100 ms contact times. J-CP was simulated by quenching Zeeman terms and reintroducing flip-flop terms of $J_{NH}$ coupling while allowing abundant water to chemically exchange with labile proton with specified exchange rates. Maximum efficiency in the absence of chemical exchange is pointed with an arrow, while maximum efficiency for different exchange rates are highlighted by black filled circles. $T_{1N}$ and $T_{2N}$ were 0.1 s; other relaxation parameters were: $T_{1w}$ = 3s, $T_{2w}$ = 0.5s, $T_{1HN}$ = 0.5 and $T_{2HN}$ = 0.2s. b) Experimental buildup of $N_X$ polarization in $^{15}$N-labeled alanine, acquired at different temperatures (hence at various chemical exchange rates). The Y-axis was normalized to the maximum signal achieved at 278 K, when the chemical exchange was the slowest and buildups the most efficient.

This Equation can be converted into two block-diagonal systems: one involving $\langle H_X \rangle$, $\langle N_X \rangle$ and $\langle 2H_Y N_Z \rangle - \langle 2H_Z N_Y \rangle$, and another one involving solely $\langle 2H_Y N_Z \rangle + \langle 2H_Z N_Y \rangle$. Analytical solutions for the 3x3 system are still complex, but relatively simple expressions can be obtained from them via Taylor expansions of the 3x3 matrix's eigenvalues and eigenvectors. In the slow exchange regime these expansions were performed with respect to the parameter $\epsilon_s = k_{HW}/J$ < 1; following standard algebra these lead to a nitrogen polarization buildup described by

$$\langle N_X \rangle(t) = \frac{\left(-k_{HW} e^{\left(-\frac{1}{2}k_{HW} - R_{eff}\right)t} + 2(R_{eff} - R_W)e^{\left(-\frac{1}{2}k_{HW} - R_{eff}\right)t} + 2k_{HW} e^{-R_W t}\right)(4\pi^2 J^2 + k_{HW}^2)}{(2k_{HW} + 4R_{eff} - 4R_W)(4\pi^2 J^2 - k_{HW}^2)}$$

$$- \frac{k_{HW}^2}{2\pi^2 J^2} e^{-R_W t} - e^{\left(-\frac{3}{4}k_{HW} - R_{eff}\right)t}\left[\frac{1}{2}\cos(\pi J t) + \frac{5k_{HW}}{8\pi J}\sin(\pi J t)\right]. \tag{8}$$

One can recognize here a leading term that builds up $^{15}N$ polarization without oscillations, and a final $cos(\pi J t)$ term of the kind introduced in Eq. (1) to describe CP's oscillatory transfer, that appears dampened by an $e^{\left(-\frac{3}{4}k_{HW} - R_{eff}\right)t}$ factor. The analytical treatment leading to the $\langle N_X \rangle(t)$ in Eq. (8) also yields expressions for the time-dependencies of the $\langle H_X \rangle$ and $\langle 2H_Y N_Z \rangle$, $\langle 2H_Z N_Y \rangle$ terms, which are presented in Supporting Information Eqs. (S1)-(S3).

To obtain analytical solutions for the CP transfer in the fast exchange regime, the exact eigenvalues and eigenvectors of the evolution generator in Eq. (7) were expanded up to second order in $\epsilon_f = J/k_{HW} < 1$. After retaining the linear and quadratic terms in $\epsilon_f$, the nitrogen polarization build up is described by

$$\langle N_X \rangle(t) = \left( \frac{2\pi^2 \epsilon_f^2}{\frac{5}{4}\pi^2 \epsilon_f^2 - 4} \right) \left[ \frac{e^{-\left(\frac{\pi^2 J^2}{2 k_{HW}} + R_{eff}\right)t}}{\pi^2 \epsilon_f^2 /2 + (R_{eff} - R_W)/k_{HW}} \right.$$
$$\left. + \frac{e^{-R_W t}\left(-1 + \frac{\pi^2}{2}\epsilon_f^2 + (R_{eff} - R_W)/k_{HW}\right)}{\frac{\pi^2}{2}\epsilon_f^2 + (R_{eff} - R_W)/k_{HW}} \right] \quad (9)$$
$$+ \left( \frac{2\pi^2 \epsilon_f^2}{\frac{5}{4}\pi^2 \epsilon_f^2 - 4} \right) \frac{\pi \epsilon_f}{2\sqrt{2}} e^{\left(\frac{\pi^2 J^2}{4 k_{HW}} - k_{HW} - R_{eff}\right)t} \sin[\pi Jt/\sqrt{2}]$$

Expressions for the other expectation values are given in the Supporting Information, Eqs. (S4)-(S6).

Figure 3 compares the analytical predictions of Eqs. (8) and (9), with numerical simulations for the evolution of $\langle N_X \rangle$ upon applying CP at infinite $\omega_1$ fields, for different exchange rates. Supporting Figures S2-S5 show similar comparisons for the $\langle H_X \rangle$ and $\langle 2H_Y N_Z \rangle$ terms. For the regimes where the Taylor expansions hold, the analytical solutions are in excellent agreement with the numerical simulations. In the slow-exchange regime theory matches well numerical simulations up to exchange rates of ~2*J*, while for the fast-exchange regime, Eq. (9) and its Eqs. (S4-S6) counterparts provide accurate solution for the evolution of the coherences from exchange rates of ≥4*J*. It follows that there is an intermediate, relatively narrow region of exchange with respect to *J*, where our analytical solutions do not predict well the spin evolution.

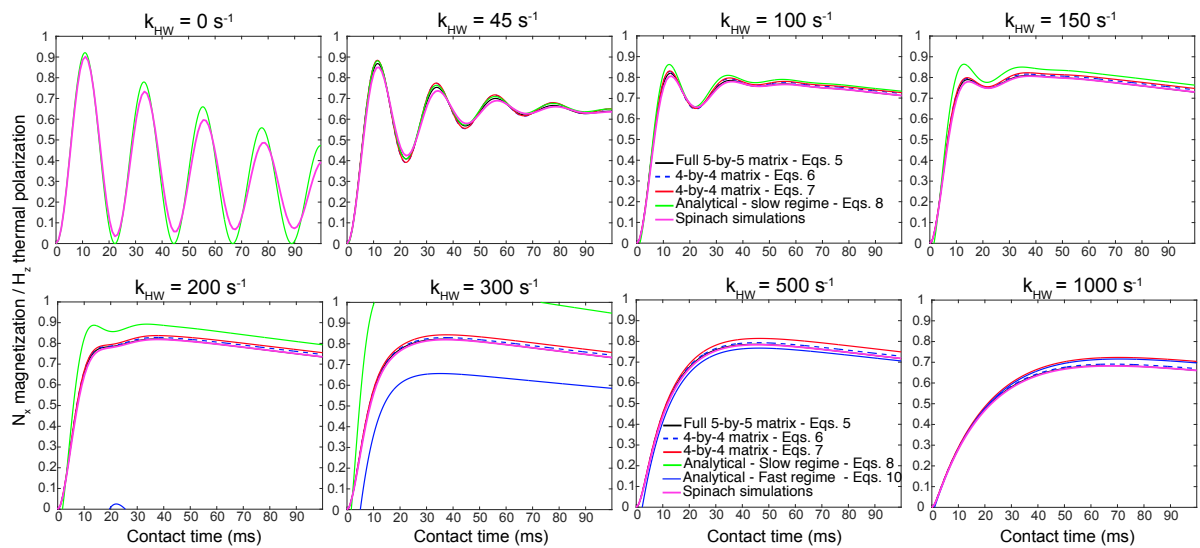

**Figure 3.** Numerically vs analytical solutions for the $\langle N_X \rangle$ arising from J-CP as a function of contact time, for various exchange rates. Curves in black result from solving numerically the full 5x5 system in Eq. (5), which always matches the Liouville-von Neumann predictions arising from Spinach (red curves). Numerical solutions of the 4x4 system in Eq. (5) (blue dashed curve) preserve this good agreement, meaning that this simpler system can also be used to reliably simulate CP in exchanging systems. Curves shown in green and blue were calculated using the analytical solutions given in Eqs. (8) and (9) valid for slow and fast exchange regimes respectively. The former provides good matches with numerical simulations for exchange rates up to 200 s$^{-1}$ while the latter works well for $k_{HW} \geq 400$ s$^{-1}$, defining an intermediate range where numerical solutions are necessary.

Figure 4a explores how chemical exchange rates influence both optimal buildup times and maximum $N_X$ amplitudes in J-CP. These dependences on chemical exchange can be divided into three parts: a slow-exchange regime where $k_{HW} \leq J$ and CP dynamics is dominated by J-coupling oscillations; an intermediate range between $J \leq k_{HW} \leq 3J$ where chemical exchange starts smearing the oscillatory buildup; and a $k_{HW} \geq 3J$ regime where oscillations have been suppressed by chemical exchange. Each regime has its own optimal CP time ($\tau_{CP}$) and efficiency behavior. In the slow exchange regime, the optimal $\tau_{CP} \approx 1/J$ and efficiency is close to maximal; the intermediate regime is fairly forgiving, with ca. 0.8 of the maximal $\langle N_X \rangle$ happening in the $1/J \leq \tau_{CP} \leq 4/J$ range; the third region shows a monotonic decay of efficiency and a concomitant increase in the optimal $\tau_{CP}$, up to a time where $R_{1\rho}^N$ becomes the main limiting factor. Figure 4b summarizes another aspect of this showing that, if relaxation times are sufficiently long, optimal $\tau_{CP}$s (blue) will polarize almost to the full $\gamma_H/\gamma_N$ extent the $^{15}$N – even for very fast exchange rates. These $\tau_{CP}$s will be considerably longer than the exchange-free theoretically optimal $1/J_{NH}$ value (Fig. 4b, in orange); Supporting Information Section 3 discusses further aspects associated with these optimal J-CP conditions under chemical exchange, including extensions of these analyses to $I_nS$ multi-spin systems involving several $^1$Hs. It follows from these calculations that in principle, as in solid state NMR, solution NMR polarization transfers approaching a $\gamma_H/\gamma_N$ maximal gain can also be reached for sufficiently long $T_{1\rho}$ relaxation times and strong, ideal spin-locking conditions–even when $k_{HW} > 20J$.

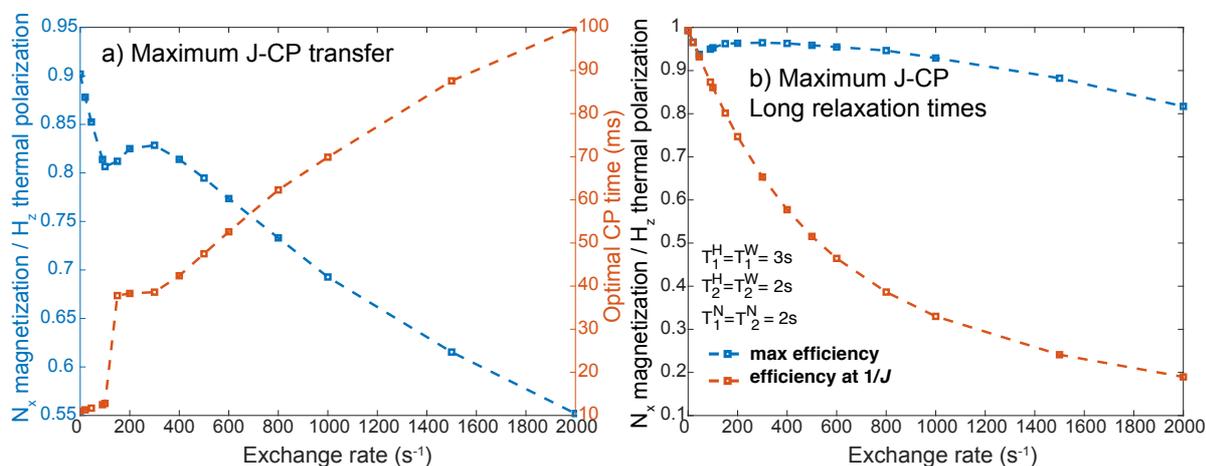

**Figure 4.** a) Reciprocal relation between the maximum $\langle N_X \rangle$ amplitude and CP contact time needed for the corresponding amplitudes to be reached. b) Maximum CP efficiency at the exchange-free $1/J_{NH}$ condition, vs at the long CP contact times that favor rapidly exchanging systems. Null relaxation rates were considered for these buildups, that therefore represent only the influence of chemical exchange on the transfer efficiency. Simulation parameters/details were the same as used in Figure 2 except for b) where relaxation times are specified in the legend; sudden inflections in the curves arise from oscillatory→monotonic buildup transitions, as pointed by the black dots in Figure 2.

Other parameters that can influence J-CP efficiency are the nutation field $\omega_1$ and the irradiation frequency offsets in the $^1$H and $^{15}$N channels. These frequencies dictate the effective field $\sqrt{\omega_1^2 + \Omega^2}$ that the spins experience; so far, they were assumed infinite ($\omega_1$) and zero ($\Omega$), respectively, apart for the potential $R_{1\rho}$ influence in Eq. (4). This is not easy to justify, given that the frequency separation between labile and water protons at high fields will be similar to the achievable spin-lock nutation field. Figure 5 shows how these factors end up balancing one another, with numerical predictions on how $\langle N_X \rangle$'s maximum amplitude will vary with respect to $\omega_1$ and $^1$H offset (given for simplicity in chemical shift ppm), for two different exchange rates. Also plotted are experimental data sets measured as a function of these two parameters. The transfer always ends up most efficient when the spin-lock is resonant with the labile proton (chosen for this case at 8 ppm), and benefitting from the highest possible $\omega_1$. For the $\omega_1$s used in the experiments below, these sufficed to effectively spin-lock labile protein's H$^N$ and H$^{water}$, up to the fields of 14 T (Figure 5c). Strong enough fields, however, are not achievable on most cryogenically-cooled NMR probes when dealing with imino sites of RNA/DNAs, resonating upward of 10 ppm.

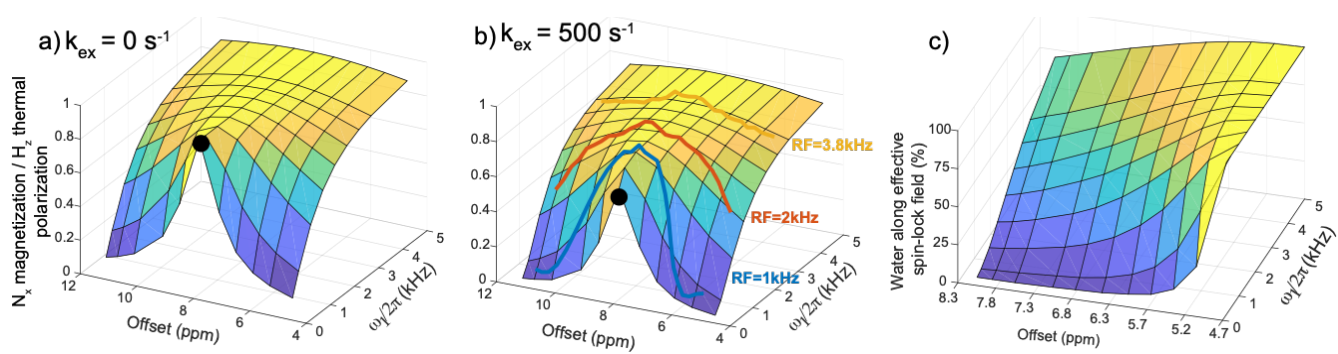

**Figure 5.** Simulated effect of CP nutation field ($\omega_1/2\pi$, in Hz) as well as offset of spin-lock on proton is examined by following optimal CP transfer for the case of (a) no chemical exchange and (b) $k_{HW} = 500\ s^{-1}$. Curves plotted over the simulated surfaces represent experimental data obtained using alanine sample at pH 5.1 at 298 K and 500 MHz magnetic field. (c) Percentage of water spin-locked along effective field after optimal $\tau_{CP}$s, for different irradiation offsets and nutation fields. Notice the small portion of water that remains spin-locked for weak irradiation fields placed on the amine/amide region, explaining why J-CP becomes very inefficient in such cases when fast exchange sets in.

Figure 6 illustrates experimentally the advantages of J-CP-based polarization transfer over INEPT-based methods, for detecting side chain amino groups in proteins. The $^{15}$Ns in these groups are notoriously hard targets to polarize yet can be important sites for inter-residue interactions. As illustrated in Figure 6, J-CP can open up an opportunity to study these at a range of relevant temperatures.

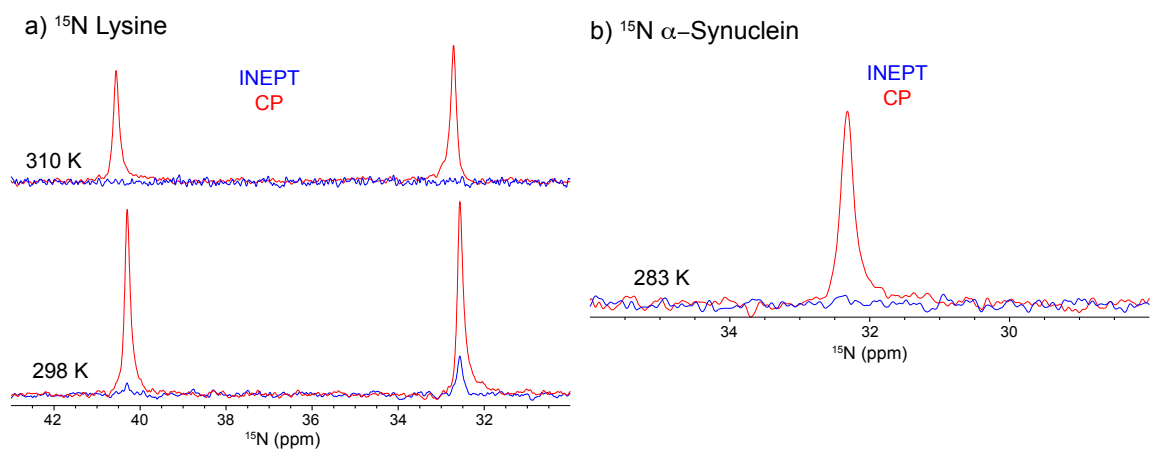

**Figure 6.** a) Detecting the two amino groups in Lysine (pH 6) at 298 and 310 K. Optimal J-CP at 298 K was $\tau_{CP}$ = 28 ms, while at 310 K $\tau_{CP}$ = 36 ms due to faster chemical exchange, both acquired using DIPSI-1 with $\omega_1/2\pi = 3500\ Hz$. b) Detecting lysine's side-chain amino groups in α-Synuclein (0.4 mM) using 40 ms of DIPSI-2 with $\omega_1/2\pi = 3500\ Hz$ at 283 K. Notice INEPT's weak/absent signals due to very fast chemical exchanges even with shorter ($\tau = 10\ ms < 1/4J$) polarization transfer times.

## Extending the $H^{water} \rightarrow H^N \rightarrow {}^{15}N$ ... polarization relay: Looped J-CP

We have recently described the benefits of relying on repeated projective measurements and related concepts, in order to improve homonuclear polarization transfers involving labile and non-labile $^1$Hs.[37,38,49] As described in such studies, suitable manipulations could then enable a $H^{water} \rightarrow H^{labile} \rightarrow H^{non\text{-}labile}$ conveyor of polarization, enhancing the latter process by a CEST-like $k_{HW}*T_1^{non\text{-}labile}$ factor. In view of this, it is interesting to explore whether looping the J-CP in a similar fashion, could also help drive the system closer to the $\gamma_H/\gamma_N$ limit. Figure 7 examines this with the aid of simulations, for conventional and looped J-CP schemes proceeding at two different chemical exchange rates. To rely on the longer longitudinal $T_{1N}$ relaxation, the looped J-CP scheme assumed ideal storage and re-excitation pulses flanking the CP periods, which were interleaved with delays $t_{ex}$ to allow $H^{water}$ to repolarize the labile $H^N$ via chemical exchange. These simulations show that, indeed, looping these blocks yields a constant buildup of the $^{15}$N magnetization. This buildup relies on a nearly complete repolarization of the $H^N$ by the solvent $^1$H, in each of the loops. Calculations also show, however, that the advantages brought about by looping are marginal. It follows that the repolarization of H$^N$ happening spontaneously by chemical exchange for as long as the H$^{water}$ and H$^N$ magnetizations are suitably spin-locked, makes up for the deleterious effects of faster chemical exchange rates on J-CP. One can then simply utilize single, long $\tau_{CP}$ to impart efficient $^{15}$N buildups: an efficient spin-lock effectively leads to a relayed $H_x^{water} \rightarrow H_x^N \rightarrow N_x$ polarization transfer process, reminiscent of the buildups in saturation-based homonuclear polarization transfer experiments in exchanging systems.[49,50]

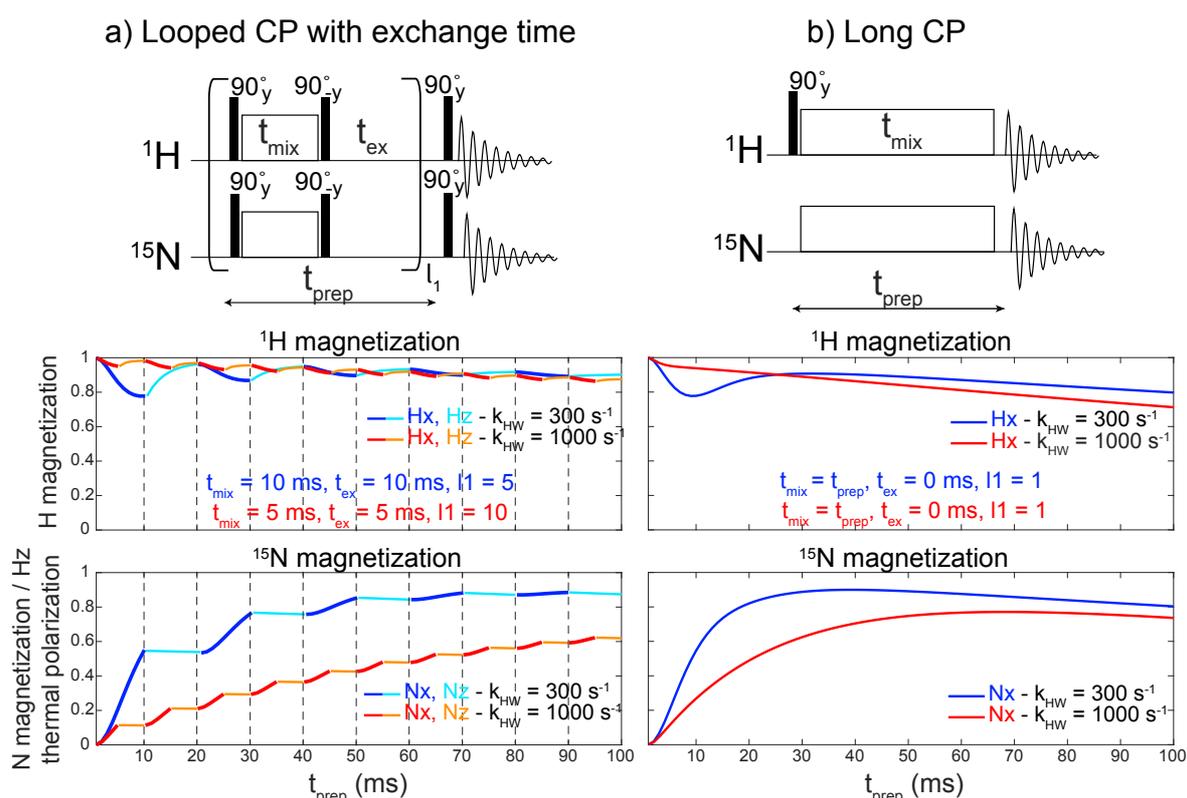

**Figure 7.** Comparing J-CP strategies to perform a $^1H \rightarrow {}^{15}N$ transfer, plotted as a function of total preparation time ($t_{prep}$) at two different exchange rates. a) Looped CP scheme involving multiple contact ($t_{mix}$) periods spaced by $t_{ex}$ exchange periods for water-driven $H^{water} \rightarrow H^N$ repolarization. Since part of the time the $^{15}$N magnetization is stored along the z-axis (to ensure use of a longer $T_{1N}$ memory time), magnetizations contain interleaved $N_X$ and $N_Z$ components (shown in different color). Ideal 90° delta-pulses were assumed. b) Conventional J-CP element, illustrate a very similar performance as the looped procedure. Simulation parameters/details were the same as used in Figure 2.

With the advantage of J-CP for these two-spin polarization transfers in exchanging systems unclear, attention was focused on triple-resonance scenarios involving transfers between the labile proton, its bound nitrogen, and a carbon *J*-coupled to the latter. Such cases often arise in NMR experiments, and are part of widely used sequences such as the HNCO and HNCA correlations.[51] Although these experiments usually transfer magnetizations between nuclei by multiple INEPT-like steps, the previous paragraphs suggest that in exchanging systems these steps might be better handled by J-CP. This is the basis of the concatenated CP (CCP) proposal by Zuiderweg and co-workers,[20] who showed that a combination of simultaneous H→$^{15}$N→$^{13}$C triple-resonance CP followed by an $^{15}$N→$^{13}$C double-resonance J-CP to account for the $J_{CN}$<$J_{NH}$ condition, could provide competitive H$^N$ → N → C counterparts to INEPT-based transfers. For the non-exchanging systems that Zuiderweg *et al* analyzed, these advantages were compared to back-to-back INEPT manipulations in terms of robustness to transverse relaxation losses. Since, as discussed in the preceding Section, J-CP can also improve polarization transfers between labile protons and nitrogen, this CCP approach could also be advantageous for fast-exchanging H$^N$s cases, such as those arising in intrinsically disordered proteins, protein sidechains, or nucleic acids.

Figure 8 examines the usefulness of J-CCP for performing H$^N$ → N$^H$ → C transfers in labile systems, by comparing the carbonyl regions in two $^{13}$C-detected 2D $^{15}$N-$^{13}$C spectra of PhoA4 –an unstructured protein whose amides are undergoing fast chemical exchange with solvent.[52] This protein was examined at two temperatures and using two different polarization transfer schemes: one based on double H→N and N→C refocused INEPT transfers; the other based on concatenated J-CP elements. The 40 ˚C spectra confirm the much higher sensitivity of the J-CCP experiment over its INEPT counterpart. Nevertheless, when the 40 ˚C CCP data are compared to the same experiment at 30 °C (Figure 8c), it becomes clear that a number of peaks are missing in the former. This is because, for this relatively fast relaxing system, the efficiency of CP is also reduced with faster chemical exchange.

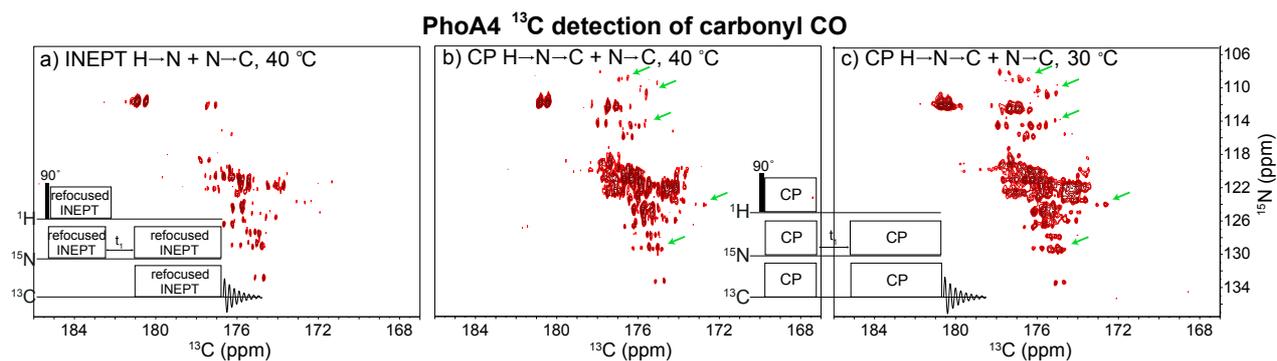

**Figure 8.** $^{13}$C-detected $^{15}$N-$^{13}$C correlations for 2.2 mM PhoA4 sample recorded using: a) double refocused INEPT transfer at 40 °C; b) a CCP at 40 °C; c) idem at 30 °C. The benefits of using J-CP are obvious upon comparing the peaks in a) and b), whose contours were done at the same level vs maximum. Still, the spectrum in c) shows some resonances of fast-exchanging amides (highlighted with green arrows) that are absent in b). $^{13}$C doublets appear along F2 since no provisions to suppress $^{13}$CO-$^{13}$Cα J-couplings were applied.

To explore whether these deleterious effects of chemical exchange in triple-resonance CCP transfers can be eliminated, we explored an adaptation of the looped repolarization concepts introduced earlier. The resulting Looped Concatenated CP (L-CCP) experiments seek to improve the $H^N \rightarrow N^H \rightarrow C$ transfer efficiency by introducing a storage period were $H^{water}$ repolarizes, by exchange, the $H^N$. As CCP, this triple-resonance CP is performed using two consecutive contacts that account for the different $\tau_{CP}$s demanded by $J_{NH} > J_{NC}$; to enable a repolarization of the $H^N$s during the second, longer CP, all steps are flanked by 90° storage/recall pulses that preserve longitudinal magnetizations. Figure 9 shows the sequence emerging from such proposals, together with numerical simulations showing how the different spins' polarizations change over the course of this repeated CCP. The initial CP puts in contact all three spin reservoirs simultaneously; during this period $^1$H polarization transfers to the $^{15}$N, whose polarization grows up to a level allowed by solvent exchanges; also the $^{13}$C gets somewhat polarized. Following 90° pulses that store back all the states along the *z*-axis (particularly the $^1$Hs, which include both $H^N$ and $H^{water}$), the $^{15}$N passes its polarization to its bonded $^{13}$C in a second J-CP, while $H^N$ recovers its original polarization thanks to fast exchanges with the solvent. This single $H^N \rightarrow N^H \rightarrow C$ transfer is of limited efficiency owing to the exchange-driven averaging of the $^1$H-$^{15}$N J-CP (Fig. 1); L-CCP compensates for this by repeatedly looping the $H^N \rightarrow N^H \rightarrow C$ buildup, while relying on the $^{13}$C's longer lifetimes. Figure 9b illustrates the behavior arising from this procedure on the three-spin system throughout multiple loops, as simulated using Eq. (2) while in the presence of chemical exchanges with a water reservoir (represented by 200 protons). Notice how, regardless of the chemical exchange rates, the multiple repetitions eventually polarize to the maximum limit given by the $^1$H thermal polarization both the $^{15}$N and the $^{13}$C, by repeatedly drawing spin polarization from the solvent. This is a procedure that is reminiscent of algorithmic cooling,[53,54] yet relying on the large polarization in the H$^{water}$ reservoir. Notice as well that given the mutual exchanges involved in the J-CP process, the $^{15}$N enhances the $^{13}$C only in odd-numbered looping events: in even-numbered events the Hartmann-Hahn contact actually enhances the $^{15}$N at expense of the latest value reached by the $^{13}$C polarization. This in turn decreases the amount of polarization that the $^{15}$N will draw when put back in contact with the $H^N$ reservoir, which thus remains more and more polarized throughout the events as the CCPs are looped. As mentioned, all species end up eventually near the fully polarized $^1$H thermal values.

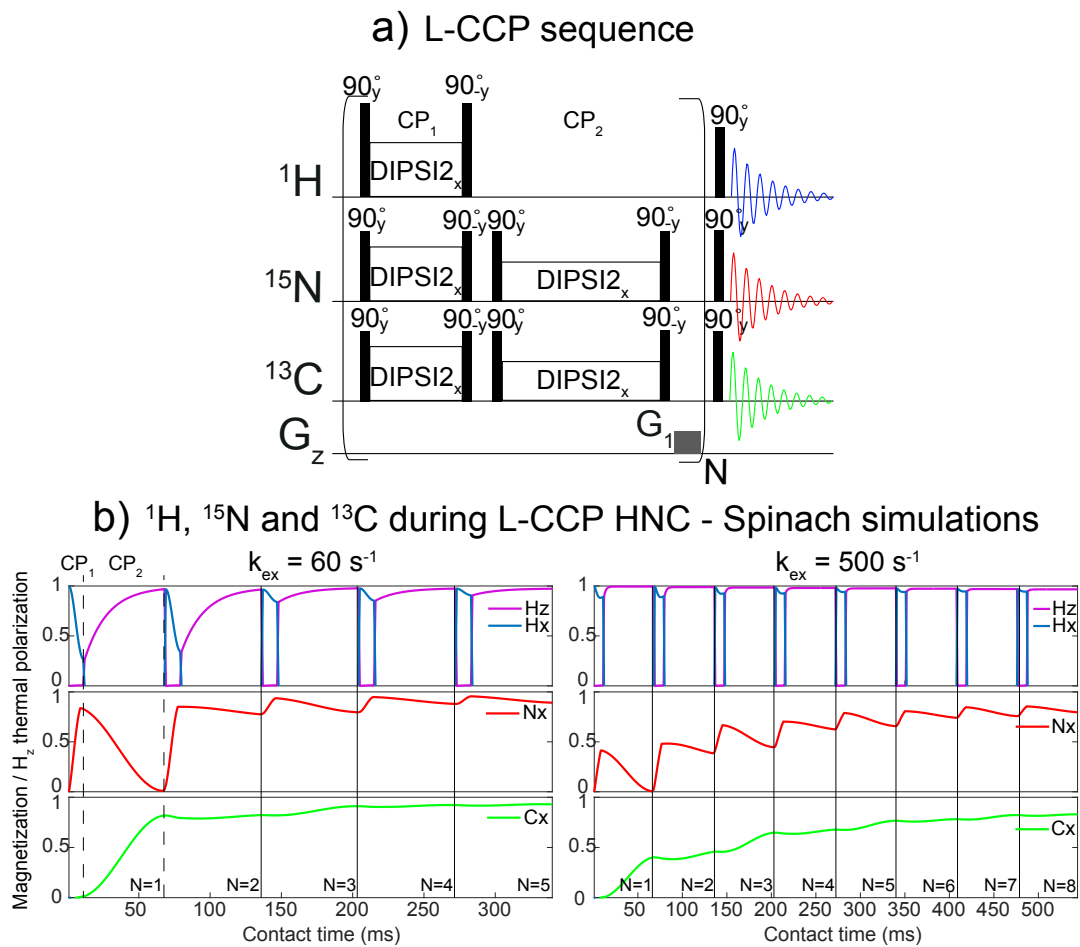

**Figure 9.** a) Looped concatenated CP (L-CCP) pulse sequence designed to improve $H^N \rightarrow N^H \rightarrow C$ transfers. The 90°$_{-y}$/90°$_y$ pairs of pulses applied on the $^{15}$N and $^{13}$C in-between the two DIPSI2 contact times can be obviated, but they are here shown to emphasize the projective nature of the procedures. b) Numerical simulations showing how magnetizations for the a three-spin $H^N$, $^{15}$N, $^{13}$C system evolve, when immersed in a bath of 200 additional water $^1$Hs chemical exchanging with $H^N$, Magnetization components are shown for two exchange rates, confirming in both cases the advantages of L-CCP for enhancing the $^{13}$C polarization. $J_{NH}$ coupling was set to 90 Hz, $J_{NC}$ was 16 Hz. All magnetizations are normalized to a maximum $^1$H longitudinal thermal polarization of one.

Figure 10a shows a validation of such theoretical expectations, based on detections of urea's carbon signal at pH=3 and at three different temperatures; i.e., at three different exchange rates with water. In all these cases fast exchanges average out the $J_{HN}$ = 90 Hz coupling, depriving INEPT-based transfers from usefulness. J-CCP works markedly better, but its efficiency also decreases with temperature, as evidenced by the progressively smaller $^{13}$C levels that can be reached (filled circles). The L-CCP scheme, by contrast, reinstates the efficiency, delivering $^{13}$C enhancements that are close to CP's theoretical $\gamma_H/\gamma_C$ = 4 maximum for all cases. Notice L-CCP's ~2-fold enhancement compared to CCP, when measured at 38 °C. Further experimental evidence is shown in Figure 10b, with carbonyl-based detection of two unstructured proteins at different temperatures. Once again, CP-based transfers are always more efficient than INEPT-based counterparts, while L-CCP always provides the highest enhancements. L-CCP-derived gains here, however, are lower than for urea; we ascribe this to the shorter $T_{1\rho}$ and $T_1$ values of the $^{15}$N and $^{13}$C in these larger biomolecules, and to the slower chemical exchange that amide protons in these proteins experience compared to urea. Nevertheless, there are multiple peaks (labeled with yellow arrows in Fig. 10b) that are significantly enhanced (>2x) when compared to the CCP acquisition, most likely because they originate from faster exchanging amide protons.

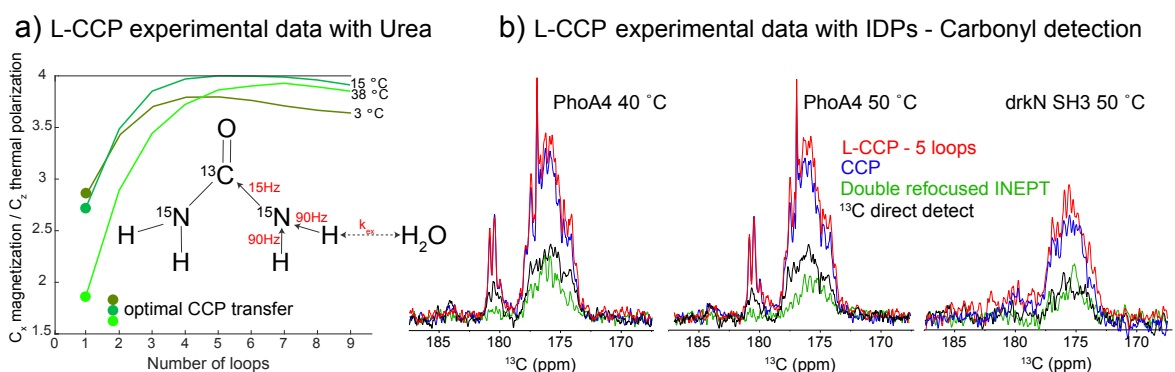

**Figure 10.** a) Experimental L-CCP results measured for urea's carbonyl $^{13}$C in H$_2$O at three different temperatures as a function of L-CCP looping. Starting point of buildups market with filled circles correspond to optimal CCP transfer at given temperature. b) Carbonyl spectra of unstructured proteins showing the superiority of CCP and L-CCP experiments comparing to INEPT or direct detect $^{13}$C.

## Conclusions and Outlook

This study revisited the nature of and proposed improvements for heteronuclear polarization transfers by J-CP, when involving rapidly exchanging protons. The superiority of J-CP over INEPT-based transfers was examined, as were the optimal conditions for J-CP as a function of contact time and exchange rates. Liouville-space simulations were used as starting point for these examinations, which showed a transition in the $^1$H→$^{15}$N polarization from an oscillatory to an exponential buildup, as the rate of exchange for the labile H$^N$ increased. This transition, reminiscent of similar transitions observed for CP in the solid state, was not associated with a dramatic drop in the net $^1$H→$^{15}$N polarization transfer efficiency, which could still approach maximal $\gamma_H/\gamma_N$ values if spin-lock relaxation times $T_{1\rho}$ remained sufficiently long. These numerical predictions could be accurately reproduced by an approximate analytical treatment, shown valid in the slow ($k_{HW}{\leq}2J$) and fast ($k_{HW}{\geq}4J$) exchange regimes. It also matched well experiments performed on model exchanging systems. Additional insight was obtained when examining this J-CP process as a function of offset and spin-locking field, where it was found that J-CP's efficiency could only be maintained for strong B$_1$ fields, capable of keeping both H$^{water}$ and H$^N$ spin-locked for prolonged periods despite the chemical exchange. To further the efficiency of these heteronuclear transfers, schemes based on looping a module involving a shorter J-CP transfer followed by a chemical exchange period, were assessed. Because of the relative efficiency of the original $^1$H→$^{15}$N J-CP process, this looped-CP strategy yielded modest improvements for two-spin processes. However, more substantial improvements were found in triple resonance experiments based on the concatenated CP approach, when a $^{13}$C is viewed as the final reservoir of the solvent-derived polarization. In the ensuing Looped-CCP experiments, the $^1$H-$^{15}$N spin pair acts as a sort of "conveyor" of polarization, between the solvent $^1$Hs and the $^{13}$C. Polarization transfers in this chain are not trivial, as CP is not a unidirectional process, but rather a polarization exchange mechanism; as a result, $^{13}$C polarization only increased significantly in odd-numbered segments of the looping. Still, both simulations and experiments showed that L-CCP could be used for polarizing $^{13}$C to nearly their $\gamma_H/\gamma_C$ maximal limit. In fact, additional Liouville-space simulations show that the L-CCP strategy can be used for facilitating the simultaneous polarization of CO and C$_\alpha$ carbons bonded to the $^{15}$N (Supporting Information Fig. S9). Polarized HNCO and HNCA 3D acquisitions –as well as their heteronuclear-detected 2D counterparts (H)NCO and (H)NCA[55]– could thus be feasible in a single experiment. Simulations indicate that L-CCP could also be beneficial for HN(CO)CA and HNCACB experiments. Considering that high proton and nitrogen polarizations are also obtained at the end of such looping, this could potentially be applied to detect various simultaneous experiments with multiple receivers.[56–58] Further efforts are in progress to explore potential applications of these triple resonance concepts on different systems –particularly disordered proteins and nucleic acids involving labile, chemically exchanging protons.

## Experimental section

### Sample preparation

$^{15}$N-labeled alanine was purchased from Cambridge Isotope laboratories and prepared at 100 mM and different pH values, ranging from 2.1 to 7. $^{15}$N Lysine was purchased from Silantes and was prepared at 100 mM and pH 6 in Sodium Phosphate buffer. $^{15}$N/$^{13}$C Urea was purchased from CIL and was prepared at 300 mM and pH 3. All pH values were adjusted using concentrated HCl and NaOH. 2.2 mM PhoA4, an unfolded protein fragment of Alkaline Phosphatase (PhoA) from *E. coli*, and 2 mM drkN SH3, were prepared in 50 mM HEPES buffer at pH 7.5 with 50 mM KCl.[52] α–Synuclein[59] was prepared at 0.4 mM concentration in 20 mM Sodium Phosphate buffer with 150 mM NaCl at pH 7.4. All samples were prepared in approximately H$_2$O:D$_2$O (90%:10%) and sodium azide was added to them as an anti-microbial agent at a final concentration of 0.02–0.05% (w/v).

### NMR experiments

NMR experiments presented in Figures 2b, 8 and 10 were conducted on an 11.7 T (500 MHz) Magnex magnet (Abingdon, UK) with a Varian iNova console (Palo Alto, USA) equipped with a double resonance HX Varian 5 mm probe with *z*-gradients. Results shown in Figure 1b were acquired on a 23.5 T Bruker magnet (1000 MHz) run with a Bruker Avance Neo and TCI cryoprobe. Data presented in Figure 5 were acquired on a 14.1 T Bruker magnet (600 MHz) equipped with an Avance III console and TCI Prodigy probe. Spectra shown in Figure 6 were obtained on a 500 MHz Magnex magnet run by a Bruker Avance Neo using a TCI Prodigy probe.

$^{15}$N INEPT experiments were acquired using standard Bruker sequence ineptrd with J-coupling set to 95 Hz and cnst11 = 6. Triple resonance HNC INEPT experiments were acquired using home-written double-refocused INEPT sequence where τ, J-evolution, delays for HN and NC transfers were optimized for maximum signal. All cross-polarization experiments were obtained using custom-written sequences involving DIPSI-2[28] spin-lock (proved to be most robust with respect to chemical exchange and offset) except for Figure 6 and Figure 2b where DIPSI-1[60] was used due to shorter cycle to allow for better sampling of experimental curve. Nutation fields of 3-4 kHz for HN transfers and 1.5 kHz for NC transfers were used throughout experiments and did not cause any significant sample heating or probe arcing on our RT and cryoprobes. However, in order to avoid a reduction in CP efficiency, nutation fields had to be

kepr <3.5 kHz when using Prodigy TCI probes, where the limiting factor is the maximum power that can be deposited on $^{15}$N RF channel. As in the simulations, the offset was always kept on resonance with the targeted proton, nitrogen and carbon resonances.

**Simulation parameters**

All Spinach simulations were performed at 298 K and 11.7 T magnetic field using $J_{NH}$ = 90 Hz. For $T_{1\rho}$ calculations, water chemical shift was set to be at 4.7 ppm (0 Hz), labile proton chemical shift at 8 ppm (1650 Hz) representing H$^N$ amide or amino proton with a proton offset set at 1500 Hz, almost on resonance with labile protons. $\omega_1/2\pi$ was taken as 3500 Hz (prolonged CP contact times are achievable with this nutation field at most room-temperature and cryo probes) unless otherwise indicated. Different $T_1$, $T_2$ and $k_{HW}$ values influencing $T_{1\rho}$ were tested in simulations and indicated in figure captions. See Supporting Information 4 for a discussion on how the size of the water reservoir was accounted for.

**Acknowledgements**

This work would have been impossible without the technical assistance of the late Koby Zibzener. We are grateful to Dr. Rina Rosenzweig and Prof. Philipp Selenko (Weizmann Institute) for the $^{15}$N$^{13}$C-labeled drkN-SH3, the PhoA4 and a-Synuclein samples, respectively. This work was supported by the EU Horizon 2020 program (FET-OPEN Grant 828946, PATHOS), Israel Science Foundation Grant 965/18, the Weizmann-UK Joint Research Programme, and the Perlman Family Foundation. MGC and JK were supported by Weizmann Faculty of Chemistry Dean Fellowships; JK is also supported by the Foreign Postdoctoral Fellowship Program of the Israel Academy of Sciences and Humanities. LF holds the Bertha and Isadore Gudelsky Professorial Chair and Heads the Clore Institute for High-Field Magnetic Resonance Imaging and Spectroscopy, whose support is acknowledged.

# References

(1) Aue, W. P.; Bartholdi, E.; Ernst, R. R. Two-Dimensional Spectroscopy. Application to Nuclear Magnetic Resonance. *J. Chem. Phys.* **1976**, *64* (5), 2229–2246. https://doi.org/10.1063/1.432450.

(2) Nagayama, K.; Wtithrich, K.; Bachmann, P.; Ernst, R. R. Two-Dimensional NMR Spectroscopy. **1977**, 3.

(3) Bodenhausen, G.; Freeman, R. Correlation of Proton and Carbon-13 Nmr Spectra by Heteronuclear Two-Dimensional Spectroscopy. *J. Magn. Reson.* **1977**, *28* (3), 471–476. https://doi.org/10.1016/0022-2364(77)90289-X.

(4) Maudsley, A. A.; Ernst, R. R. Indirect Detection of Magnetic Resonance by Heteronuclear Two-Dimensional Spectroscopy. *Chem. Phys. Lett.* **1977**, *50* (3), 368–372. https://doi.org/10.1016/0009-2614(77)80345-X.

(5) Morris, G. a.; Freeman, R. Enhancement of Nuclear Magnetic Resonance Signals by Polarization Transfer. *J. Am. Chem. Soc.* **1979**, *233* (2), 760–762. https://doi.org/Doi 10.1021/Ja00497a058.

(6) Mueller, L. Sensitivity Enhanced Detection of Weak Nuclei Using Heteronuclear Multiple Quantum Coherence. *J. Am. Chem. Soc.* **1979**, *101* (16), 4481–4484. https://doi.org/10.1021/ja00510a007.

(7) Burum, D. P.; Ernst, R. R. Net Polarization Transfer via a J-Ordered State for Signal Enhancement of Low-Sensitivity Nuclei. *J. Magn. Reson.* **1980**, *39* (1), 163–168. https://doi.org/10.1016/0022-2364(80)90168-7.

(8) Doddrell, D. M.; Pegg, D. T.; Bendall, M. R. Distortionless Enhancement of NMR Signals by Polarization Transfer. *J. Magn. Reson.* **1982**, *48* (2), 323–327. https://doi.org/10.1016/0022-2364(82)90286-4.

(9) Bodenhausen, G.; Ruben, D. J. Natural Abundance Nitrogen-15 NMR by Enhanced Heteronuclear Spectroscopy. *Chem. Phys. Lett.* **1980**, *69* (1).

(10) Redfield, A. G. Stimulated Echo Nmr Spectra and Their Use for Heteronuclear Two-Dimensional Shift Correlation. *Chem. Phys. Lett.* **1983**, *96* (5), 537–540.

(11) Bendall, M. R.; Pegg, D. T.; Doddrell, D. M. Pulse Sequences Utilizing the Correlated Motion of Coupled Heteronuclei in the Transverse Plane of the Doubly Rotating Frame. *J. Magn. Reson.* **1983**, *117*, 81–117.

(12) Bax, A.; Griffey, R. H.; Hawkins, B. L. Correlation of Proton and Nitrogen-15 Chemical Shifts by Multiple Quantum NMR. *J. Magn. Reson.* **1983**, *55*, 301–315. https://doi.org/10.1016/0022-2364(83)90241-X.

(13) Krishnan, V. V; Rance, M. Influence of Chemical Exchange among Homonuclear Spins in Heteronuclear Coherence-Transfer Experiments in Liquids. *J. Magn. Reson. A* **1995**, *116* (1), 97–106. https://doi.org/10.1006/jmra.1995.1194.

(14) Liu, A.; Majumdar, A.; Hu, W.; Kettani, A.; Skripkin, E.; Patel, D. J. NMR Detection of N-H⋯O=C Hydrogen Bonds In13C,15N-Labeled Nucleic Acids. *J. Am. Chem. Soc.* **2000**, *122* (13), 3206–3210. https://doi.org/10.1021/ja994255s.

(15) Yuwen, T.; Skrynnikov, N. R. CP-HISQC: A Better Version of HSQC Experiment for Intrinsically Disordered Proteins under Physiological Conditions. *J. Biomol. NMR* **2014**, *58*, 175–192. https://doi.org/10.1007/s10858-014-9815-5.

(16) Bertrand, R. D.; Moniz, W. B.; Garroway, A. N.; Chingas, G. C. Carbon-13-Proton Cross-Polarization in Liquids. *J. Am. Chem. Soc.* **1978**, *100* (16), 5227–5229. https://doi.org/10.1021/ja00484a063.

(17) Chingas, G. C.; Garroway, A. N.; Moniz, W. B.; Bertrand, R. D. Adiabatic J Cross-Polarization in Liquids for Signal Enhancement in NMR. *J. Am. Chem. Soc.* **1980**, *102* (8), 2526–2528. https://doi.org/10.1021/ja00528a002.

(18) Ernst, M.; Griesinger, C.; Ernst, R. R.; Bermel, W. Optimized Heteronuclear Cross Polarization in Liquids. *Mol. Phys.* **1991**, *74* (2), 219–252. https://doi.org/10.1080/00268979100102191.

(19) Zuiderweg, E. R. P. Analysis of Multiple-Pulse-Based Heteronuclear J Cross Polarization in Liquids. *J. Magn. Reson.* **1990**, *89* (3), 533–542. https://doi.org/10.1016/0022-2364(90)90336-8.

(20) Majumdar, A.; Zuiderweg, E. R. P. Efficiencies of Double- and Triple-Resonance J Cross Polarization in Multidimensional NMR. *Journal of Magnetic Resonance, Series A*. 1995, pp 19–31. https://doi.org/10.1006/jmra.1995.1051.

(21) Hartmann, S. R.; Hahn, E. L. Nuclear Double Resonance in the Rotating Frame. *Phys. Rev.* **1962**, *128* (5), 2042–2053. https://doi.org/10.1103/PhysRev.128.2042.

(22) Keeler, J. *Understanding NMR Spectroscopy*; Wiley, 2010.

(23) Pines, A.; Gibby, M. G.; Waugh, J. S. Proton-enhanced NMR of Dilute Spins in Solids. *J. Chem. Phys.* **1973**, *59* (2), 569–590. https://doi.org/10.1063/1.1680061.

(24) Schaefer, J.; Stejskal, E. O.; Buchdahl, R. Magic-Angle 13C NMR Analysis of Motion in Solid Glassy Polymers. *Macromolecules* **1977**, *10* (2), 384–405. https://doi.org/10.1021/ma60056a031.

(25) Lopez, J.; Schneider, R.; Cantrelle, F. X.; Huvent, I.; Lippens, G. Studying Intrinsically Disordered Proteins under True In Vivo Conditions by Combined Cross-Polarization and Carbonyl-Detection NMR Spectroscopy. *Angew. Chem. - Int. Ed.* **2016**, *55* (26), 7418–7422. https://doi.org/10.1002/anie.201601850.

(26) Levitt, M. H.; Freeman, R.; Frenkiel, T. Broadband Heteronuclear Decoupling. *J. Magn. Reson. 1969* **1982**, *47* (2), 328–330. https://doi.org/10.1016/0022-2364(82)90124-X.

(27) Bax, A. D.; Davis, D. G. MLEV-17-Based Two-Dimensional Homonuclear Magnetization Transfer Spectroscopy. *J. Magn. Reson.* **1985**, *65*, 355–360.

(28) Rucker, S.; Shaka, A. J. Broadband Homonuclear Cross Polarization in 2D N.M.R. Using DIPSI-2. *Mol. Phys.* **1989**, *68* (2), 509–517. https://doi.org/10.1080/00268978900102331.

(29) Shaka, A. J.; Keeler, J.; Frenkiel, T.; Freeman, R. An Improved Sequence for Broadband Decoupling: WALTZ-16. *J. Magn. Reson.* **1983**, *52* (2), 335–338. https://doi.org/10.1016/0022-2364(83)90207-X.

(30) Bearden, D. W.; Brown, L. R. Heteronuclear Isotropic Mixing in Liquids. *Chem. Phys. Lett.* **1989**, *163* (4–5), 432–436. https://doi.org/10.1016/0009-2614(89)85163-2.

(31) Furrer, J.; Kramer, F.; Marino, J. P.; Glaser, S. J.; Luy, B. Homonuclear Hartmann-Hahn Transfer with Reduced Relaxation Losses by Use of the MOCCA-XY16 Multiple Pulse Sequence. *J. Magn. Reson.* **2004**, *166* (1), 39–46. https://doi.org/10.1016/j.jmr.2003.09.013.

(32) Brown, L. R.; Sanctuary, B. C. Hetero-TOCSY Experiments with WALTZ and DIPSI Mixing Sequences. *J. Magn. Reson.* **1991**, *91* (2), 413–421. https://doi.org/10.1016/0022-2364(91)90207-A.

(33) Parella, T. A Complete Set of Novel 2D Correlation NMR Experiments Based on Heteronuclear J-Cross Polarization. *J. Biomol. NMR* **2004**, *29* (1), 37–55. https://doi.org/10.1023/B:JNMR.0000019514.04047.8b.

(34) Nolis, P.; Parella, T. Solution-State NMR Experiments Based on Heteronuclear Cross-Polarization. *Curr. Anal. Chem.* **2007**, *3* (1), 47–68. https://doi.org/10.2174/157341107779314262.

(35) Szyperski, T.; Luginbühl, P.; Otting, G.; Güntert, P.; Wüthrich, K. Protein Dynamics Studied by Rotating Frame 15N Spin Relaxation Times. *J. Biomol. NMR* **1993**, *3* (2), 151–164. https://doi.org/10.1007/BF00178259.

(36) Sahu, S. C.; Bhuyan, A. K.; Udgaonkar, J. B.; Hosur, R. V. Backbone Dynamics of Free Barnase and Its Complex with Barstar Determined by 15N NMR Relaxation Study. *J. Biomol. NMR* **2000**, *18* (2), 107–118. https://doi.org/10.1023/A:1008310402933.

(37) Novakovic, M.; Cousin, S. F.; Jaroszewicz, M. J.; Rosenzweig, R.; Frydman, L. Looped-PROjected SpectroscopY (L-PROSY): A Simple Approach to Enhance Backbone/Sidechain Cross-Peaks in 1H NMR. *J. Magn. Reson.* **2018**, *294*, 169–180. https://doi.org/10.1016/j.jmr.2018.07.010.

(38) Novakovic, M.; Battistel, M. D.; Azurmendi, H. F.; Concilio, M.-G.; Freedberg, D. I.; Frydman, L. The Incorporation of Labile Protons into Multidimensional NMR Analyses: Glycan Structures Revisited. *J. Am. Chem. Soc.* **2021**, *143* (23), 8935–8948. https://doi.org/10.1021/jacs.1c04512.

(39) Hogben, H. J.; Krzystyniak, M.; Charnock, G. T. P.; Hore, P. J.; Kuprov, I. Spinach - A Software Library for Simulation of Spin Dynamics in Large Spin Systems. *J. Magn. Reson.* **2011**, *208* (2), 179–194. https://doi.org/10.1016/j.jmr.2010.11.008.

(40) Kuprov, I. Diagonalization-Free Implementation of Spin Relaxation Theory for Large Spin Systems. *J. Magn. Reson.* **2011**, *209* (1), 31–38. https://doi.org/10.1016/j.jmr.2010.12.004.

(41) Kuprov, I. Large-Scale NMR Simulations in Liquid State: A Tutorial. *Magn. Reson. Chem.* **2018**, *56* (6), 415–437. https://doi.org/10.1002/mrc.4660.

(42) Abragam, A.; Goldman, M. Principles of Dynamic Nuclear Polarisation. *Rep. Prog. Phys.* **1978**, *41* (3), 395.

(43) Goldman, M. *Quantum Description of High-Resolution NMR in Liquids*; Clarendon Press, Oxford, 1990. https://doi.org/10.1063/1.2810567.

(44) Cavanagh, J.; Fairbrother, W. J.; Palmer, A. G.; Rance, M.; J., S. N. *Protein NMR Spectroscopy*; Elsevier, 2007. https://doi.org/10.1016/B978-012164491-8/50005-1.

(45) Trott, O.; Palmer, A. G. R1ρ Relaxation Outside of the Fast-Exchange Limit. *J. Magn. Reson.* **2002**, *154* (1), 157–160. https://doi.org/10.1006/jmre.2001.2466.

(46) Wong, L. E.; Kim, T. H.; Rennella, E.; Vallurupalli, P.; Kay, L. E. Confronting the Invisible: Assignment of Protein 1 H N Chemical Shifts in Cases of Extreme Broadening. *J. Phys. Chem. Lett.* **2020**, *11* (9), 3384–3389. https://doi.org/10.1021/acs.jpclett.0c00747.

(47) Müller, L.; Kumar, A.; Baumann, T.; Ernst, R. R. Transient Oscillations in NMR Cross-Polarization Experiments in Solids. *Phys. Rev. Lett.* **1974**, *32* (25), 1402–1406. https://doi.org/10.1103/PhysRevLett.32.1402.

(48) Ernst, M.; Zimmermann, H.; Meier, B. H. A Simple Model for Heteronuclear Spin Decoupling in Solid-State NMR. *Chem. Phys. Lett.* **2000**, *317* (6), 581–588. https://doi.org/10.1016/S0009-2614(99)01423-2.

(49) Novakovic, M.; Kupče, Ē.; Oxenfarth, A.; Battistel, M. D.; Darón, I.; Schwalbe, H.; Frydman, L. Hadamard Magnetization Transfers Achieve Dramatic Sensitivity Enhancements in Homonuclear Multidimensional NMR Correlations of Labile Sites in Proteins , Polysaccharides and Nucleic Acids. **2020**, 1–25.

(50) Novakovic, M.; Kupče, Ē.; Scherf, T.; Oxenfarth, A.; Schnieders, R.; Grün, J. T.; Wirmer-Bartoschek, J.; Richter, C.; Schwalbe, H.; Frydman, L. Magnetization Transfer to Enhance NOE Cross-Peaks among Labile Protons: Applications to Imino–Imino Sequential Walks in SARS-CoV-2-Derived RNAs. *Angew. Chem.* **2021**, *133* (21), 11991–11998. https://doi.org/10.1002/ange.202015948.

(51) Kay, L. E.; Ikura, M.; Tschudin, R.; Bax, A. Three-Dimensional Triple-Resonance NMR Spectroscopy of Isotopically Enriched Proteins. *J. Magn. Reson.* **1990**, *89*, 496–514. https://doi.org/10.1016/j.jmr.2011.09.004.

(52) Szekely, O.; Olsen, G. L.; Novakovic, M.; Rosenzweig, R.; Frydman, L. Assessing Site-Specific Enhancements Imparted by Hyperpolarized Water in Folded and Unfolded Proteins by 2D HMQC NMR. *J. Am. Chem. Soc.* **2020**, *142* (20), 9267–9284. https://doi.org/10.1021/jacs.0c00807.

(53) Schulman, L. J.; Mor, T.; Weinstein, Y. Physical Limits of Heat-Bath Algorithmic Cooling. *Phys. Rev. Lett.* **2005**, *94* (12), 120501. https://doi.org/10.1103/PhysRevLett.94.120501.

(54) Baugh, J.; Moussa, O.; Ryan, C. A.; Nayak, A.; Laflamme, R. Experimental Implementation of Heat-Bath Algorithmic Cooling Using Solid-State Nuclear Magnetic Resonance. *Nature* **2005**, *438* (7067), 470–473. https://doi.org/10.1038/nature04272.

(55) Bermel, W.; Felli, I. C.; Kümmerle, R.; Pierattelli, R. 13C Direct-Detection Biomolecular NMR. *Concepts Magn. Reson. Part A* **2008**, *32A* (3), 183–200. https://doi.org/10.1002/cmr.a.20109.

(56) Kupče, Ē.; Kay, L. E. Parallel Acquisition of Multi-Dimensional Spectra in Protein NMR. *J. Biomol. NMR* **2012**, *54* (1), 1–7. https://doi.org/10.1007/s10858-012-9646-1.

(57) Kupče, E.; Kay, L. E.; Freeman, R. Detecting the "Afterglow" of $^{13}$C NMR in Proteins Using Multiple Receivers. *J. Am. Chem. Soc.* **2010**, *132* (51), 18008–18011. https://doi.org/10.1021/ja1080025.

(58) Kupče, Ē.; Mote, K. R.; Madhu, P. K. Experiments with Direct Detection of Multiple FIDs. *J. Magn. Reson.* **2019**, *304*, 16–34. https://doi.org/10.1016/j.jmr.2019.04.018.

(59) Theillet, F.-X.; Binolfi, A.; Bekei, B.; Martorana, A.; Rose, H. M.; Stuiver, M.; Verzini, S.; Lorenz, D.; van Rossum, M.; Goldfarb, D.; Selenko, P. Structural Disorder of Monomeric α-Synuclein Persists in Mammalian Cells. *Nature* **2016**, *530* (7588), 45–50. https://doi.org/10.1038/nature16531.

(60) Shaka, A. J.; Lee, J.; Pines, A. Iterative Schemes for Bilinear Operators; Application to Spin Decoupling. *J. Magn. Reson.* **1988**, *77*, 274–293.

# Supporting Information for

## *Heteronuclear transfers from labile protons in biomolecular NMR: Cross Polarization, revisited*

Mihajlo Novakovic[1], Sundaresan Jayanthi[2], Adonis Lupulescu[3], Maria Grazia Concilio[1], Jihyun Kim[1], David Columbus[1], Ilya Kuprov[4], and Lucio Frydman[1*]

[1]Department of Chemical and Biological Physics, Weizmann Institute of Science, Rehovot 7610001, Israel
[2]Department of Physics, Indian Institute of Space Science and Technology, Valiamala, Thiruvananthapuram 695 547, Kerala, India
[3]Nicolae Titulescu nr. 8, Turda, Jud. Cluj, Romania
[4]School of Chemistry, University of Southampton, Southampton SO17 1BJ, UK

*Email: lucio.frydman@weizmann.ac.il

## Supporting Information 1: Build-up of $^{15}$N and depletion of labile $^{1}H_N$ polarization in J-CP subject to exchange

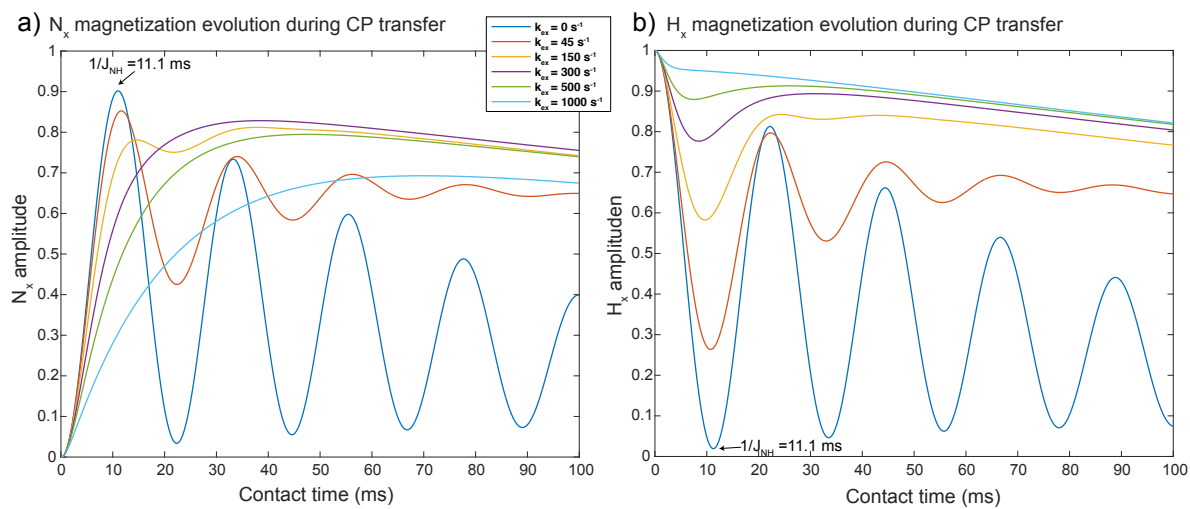

**Figure S1**. Buildup of nitrogen $N_x$ polarization upon cross-polarization for different $k_{ex}$ rates between water and labile proton. The number of water protons was chosen 200, comprising sufficiently larger spin pool than solute. To examine as well the effect of nitrogen relaxation, $T_{1N}$ and $T_{2N}$ were 0.1 s, realistic values for relaxation constants in proteins. All buildups are carried out up to 100 ms of CP contact time. Maximum efficiency in the absence of chemical exchange was pointed in the plot. Other relaxation parameters were: $T_{1w}$ = 3s, $T_{2w}$ = 0.5s, $T_1^{HN}$ = 0.5 and $T_{2H}^N$ = 0.2s.

# Supporting Information 2: Analytical solutions of J-CP under the presence of exchange for all coherences involved in the process

## 2.1 Solutions in the slow $k_{ex} < J$ exchange regime:

$$\langle N_X \rangle(t) = \frac{\left(-k_{HW}e^{\left(-\frac{1}{2}k_{HW} - R_{eff}\right)t} + 2(R_{eff} - R_W)e^{\left(-\frac{1}{2}k_{HW} - R_{eff}\right)t} + 2k_{HW}e^{-R_Wt}\right)(4\pi^2J^2 + k_{HW}^2)}{(2k_{HW} + 4R_{eff} - 4R_W)(4\pi^2J^2 - k_{HW}^2)}$$

$$- \frac{k_{HW}^2}{2\pi^2J^2}e^{-R_Wt} - e^{\left(-\frac{3}{4}k_{HW} - R_{eff}\right)t}\left[\frac{1}{2}\cos(\pi Jt) + \frac{5k_{HW}}{8\pi J}\sin(\pi Jt)\right]$$

(S1)

$$\langle H_X \rangle(t) = \frac{\left((2R_{eff} - 2R_W - k_{HW})e^{\left(-\frac{1}{2}k_{HW} - R_{eff}\right)t} + 2k_{HW}e^{-R_Wt}\right)}{(2k_{HW} + 4R_{eff} - 4R_W)}$$

$$+ \frac{k_{HW}(R_{eff} - R_W)}{2\pi^2J^2}e^{-R_Wt} + e^{\left(-\frac{3}{4}k_{HW} - R_{eff}\right)t}\left[\frac{1}{2}\cos(\pi Jt) - \frac{k_{HW}}{8\pi J}\sin(\pi Jt)\right]$$

(S2)

$\langle 2H_YN_Z \rangle(t) = -\langle 2H_ZN_Y \rangle(t)$ (this is clear from the equations of motion)

$$= -\frac{\left(k_{HW}e^{\left(-\frac{1}{2}k_{HW} - R_{eff}\right)t}(2R_{eff} - 2R_W - k_{HW}) + 2k_{HW}^2e^{-R_Wt}\right)}{4\pi J(k_{HW} + 2R_{eff} - 2R_W)}$$

$$+ \frac{k_{HW}}{4\pi J}e^{-R_Wt} - e^{\left(-\frac{3}{4}k_{HW} - R_{eff}\right)t}\left[\frac{k_{HW}}{8\pi J}\cos(\pi Jt) - \frac{1}{4}\sin(\pi Jt)\right]$$

(S3)

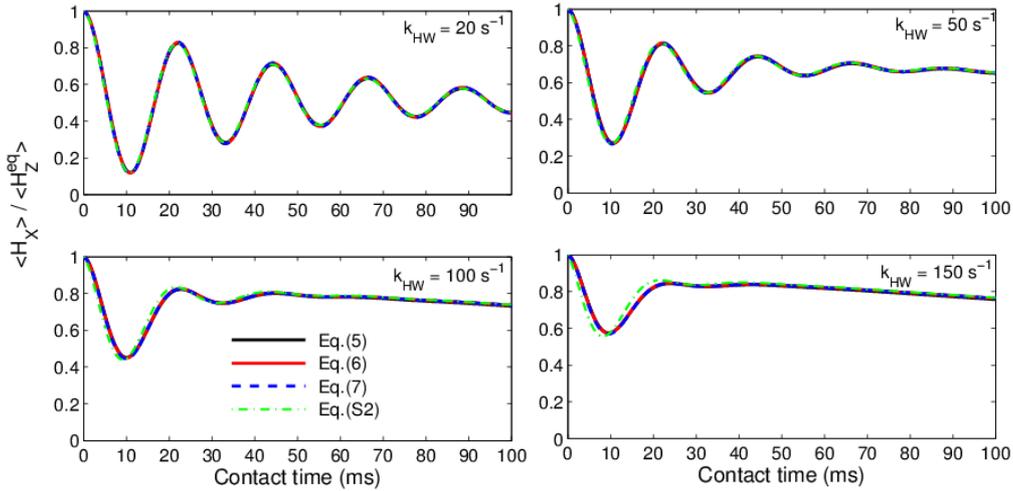

**Figure S2.** Build-up of proton magnetization $\langle H_X \rangle$ for different exchange rates in the slow exchange regime. Evolution is computed with brute force numerical simulations arising from solving Eqs. (5-7) in the main text, and compared with the analytical solution of Eq. (S2). Parameters used are $R_W = 2s^{-1}, R_H = 5\ s^{-1}, R_N = 10\ s^{-1}, J = 90\ Hz$. Initial conditions are $\langle W_X \rangle(0) = 200, \langle H_X \rangle(0) = 1, \langle N_X \rangle(0) = 0, \langle 2H_YN_Z \rangle(0) = 0, \langle 2H_ZN_Y \rangle(0) = 0$.

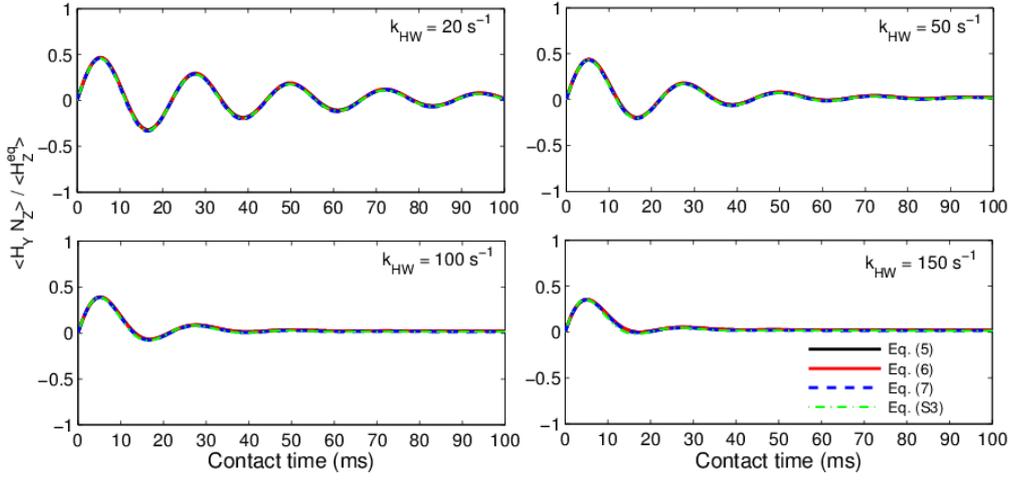

**Figure S3**. Build-up of $\langle H_Y N_Z \rangle$ for different exchange rates in the slow exchange regime. Evolution is computed with brute force numerical solutions of Eqs. (5-7) and compared with the analytical solution of Eq. (S3). Parameters used are $R_W = 2 s^{-1}, R_H = 5 s^{-1}, R_N = 10 s^{-1}, J = 90\ Hz$. Initial conditions are $\langle W_X \rangle(0) = 200, \langle H_X \rangle(0) = 1, \langle N_X \rangle(0) = 0, \langle 2 H_Y N_Z \rangle(0) = 0,\ \langle 2 H_Z N_Y \rangle(0) = 0$.

**2.2 Solutions in the fast $k_{ex} > J$ exchange regime:**

$$N_X(t) = \left(\frac{2\pi^2 \epsilon_f^2}{\frac{5}{4}\pi^2 \epsilon_f^2 - 4}\right)\left[\frac{e^{-\left(\frac{\pi^2 J^2}{2\ k_{HW}} + R_{eff}\right)t}}{\frac{\pi^2}{2}\epsilon_f^2 + (R_{eff} - R_W)/k_{HW}}\right.$$

$$\left. + \frac{e^{-R_W t}\left(-1 + \frac{\pi^2}{2}\epsilon_f^2 + (R_{eff} - R_W)/k_{HW}\right)}{\frac{\pi^2}{2}\epsilon_f^2 + (R_{eff} - R_W)/k_{HW}}\right]$$

$$+ \left(\frac{2\pi^2 \epsilon_f^2}{\frac{5}{4}\pi^2 \epsilon_f^2 - 4}\right)\frac{\pi \epsilon_f}{2\sqrt{2}}\ e^{\left(\frac{\pi^2 J^2}{4\ k_{HW}} - k_{HW} - R_{eff}\right)t}\mathrm{Sin}[\pi J t/\sqrt{2}] \quad (S4)$$

$$H_X(t) = \left(\frac{1}{4 - \frac{5}{4}\pi^2\epsilon_f^2}\right)\left[\frac{k_{HW}^2\left(4 - \frac{13}{4}\pi^2\epsilon_f^2\right)}{(R_{eff} - R_W + k_{HW})^2} e^{-R_W t}\right.$$

$$+ \left(\frac{e^{\left(\frac{\pi^2 J^2}{4 k_{HW}} - k_{HW} - R_{eff}\right)t}}{(R_{eff} - R_W + k_{HW})^2}\right)\left(k_{HW}\left(4R_{eff} - 4R_W + 2\pi^2 J\epsilon_f\right)\cos(\pi J t/\sqrt{2})\right.$$

$$+ 2\sqrt{2}\pi J k_{HW} \sin(\pi J t/\sqrt{2}))$$

$$\left.+ 2\pi^2\epsilon_f^2\left(e^{-R_W t} - e^{\left(-\frac{\pi^2 J^2}{2 k_{HW}} - R_{eff}\right)t}\right)\right] \quad (S5)$$

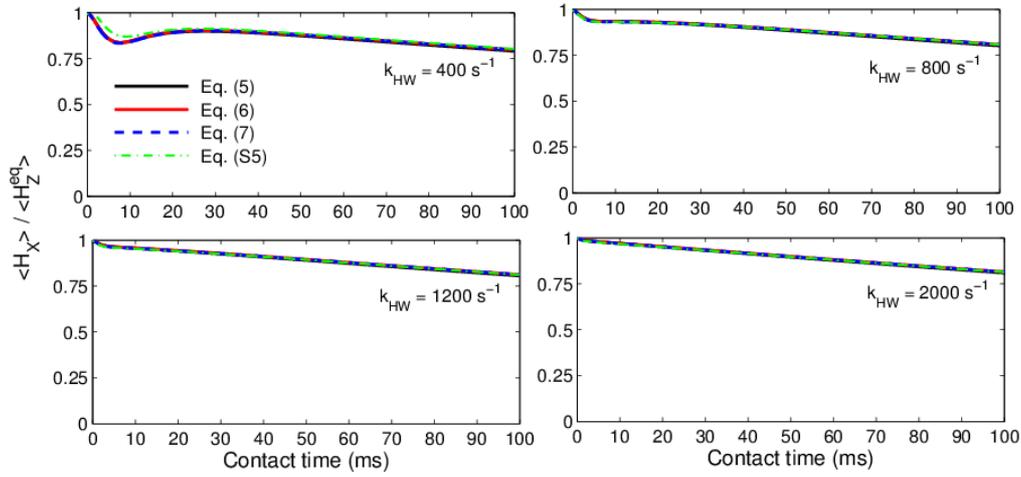

**Figure S4**. Build-up of proton magnetization $\langle H_X\rangle$ for different exchange rates in the fast exchange regime. Evolution is computed with brute force numerical solutions of Eqs. (5-7) and compared with the analytical solution of Eq. (S5). Parameters used are $R_W = 2 s^{-1}, R_H = 5 s^{-1}, R_N = 10 s^{-1}, J = 90\ Hz$. Initial conditions are $\langle W_X\rangle(0) = 200, \langle H_X\rangle(0) = 1, \langle N_X\rangle(0) = 0, \langle 2H_Y N_Z\rangle(0) = 0, \langle 2H_Z N_Y\rangle(0) = 0$.

$$\langle 2H_Y N_Z \rangle = -\langle 2H_Z N_Y \rangle$$

$$= \frac{e^{\left(\frac{\pi^2 J^2}{4\,k_{HW}} - k_{HW} - R_{eff}\right)t}}{(R_{eff} - R_W + k_{HW})^2 \left(4 - \frac{5}{4}\pi^2 \epsilon_f^2\right)} \left[-2\pi J k_{HW} \cos\left(\frac{\pi J t}{\sqrt{2}}\right) + \frac{\pi^2 J^2}{\sqrt{2}} \sin(\pi J t/\sqrt{2})\right] + \frac{2\pi J k_{HW} e^{-R_W t}}{(R_{eff} - R_W + k_{HW})^2 \left(4 - \frac{5}{4}\pi^2 \epsilon_f^2\right)}$$

$$+ \frac{\pi^3 J^2 \epsilon_f}{\left(\frac{\pi^2}{2} J^2 + R_{eff} - R_W\right)\left(4 - \frac{5}{4}\pi^2 \epsilon_f^2\right)} \left[e^{\left(-\frac{\pi^2 J^2}{2\,k_{HW}} - R_{eff}\right)t} - e^{-R_W t}\right] \quad (S6)$$

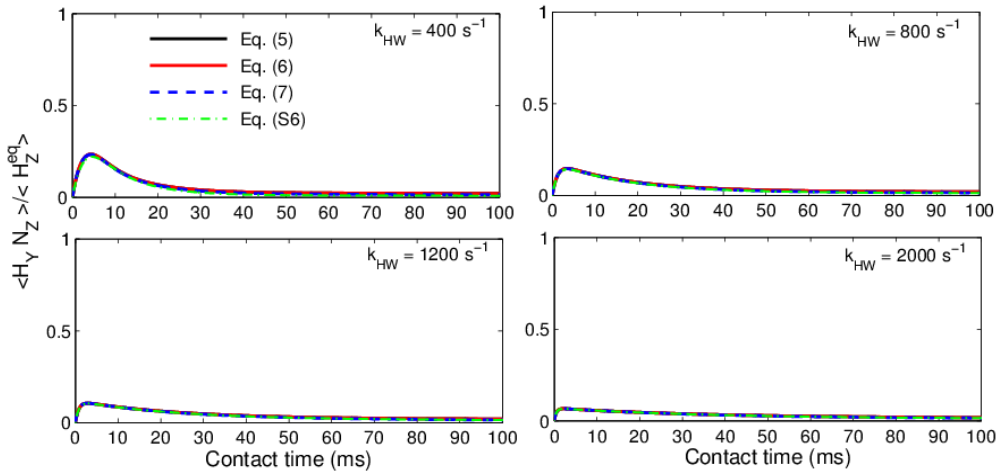

**Figure S5**. Build-up of $\langle H_Y N_Z \rangle$ for different exchange rates in the fast exchange regime. Evolution is computed with brute force numerical solutions of Eqs. (5-7) and compared with the analytical solution of Eq. (S6). Parameters used are $R_W = 2 s^{-1}, R_H = 5\ s^{-1}, R_N = 10\ s^{-1}, J = 90\ Hz$. Initial conditions are $\langle W_X \rangle(0) = 200, \langle H_X \rangle(0) = 1, \langle N_X \rangle(0) = 0, \langle 2H_Y N_Z \rangle(0) = 0,\ \langle 2H_Z N_Y \rangle(0) = 0$.

# Supporting Information 3: Additional aspects of $^1H \rightarrow {}^{15}N$ J-CP transfers under chemical exchange

Figure S6 examines the J-CP transfer performance with respect to relaxation parameters of protons and nitrogen. An optimal mixing contact was found for each case, and the behavior is plotted with respect to chemical exchange for different relaxation parameters. As expected, J-CP benefits from longer $T_1$s and $T_2$s, which make the transfer more efficient for any chemical exchange rate (in blue). Faster $^1H$ relaxation decreases J-CP's efficiency but only for higher chemical exchange rates (red), while fast $^{15}N$ relaxation (yellow) decreases the efficiency for all exchange rates. The importance of water's $T_1/T_2$ at faster exchange rates can be appreciated from curves shown in purple and green. For slow exchange rates ($k_{ex} \leq J$) the transfer is mostly influenced by the relaxation properties of exchangeable protons in the $H_x^N \rightarrow N_x^H$ process; for fast exchange rates, however, a three-way $H_x^{water} \rightarrow H_x^N \rightarrow N_x$ transfer is effectively established, whereby the relaxation properties of the $H^{water}$ define the efficiency of the J-CP. It follows that for sufficiently fast exchange rates, a full $^{15}N$ buildup will only be achieved if the water $^1H$ and the $^{15}N$ $T_{1\rho}$s are longer than the total $\tau_{CP}$ duration. This is almost always the case for water, but for large proteins the $^{15}N$ $T_{1\rho}$ will be shorter and transfer efficiency will plateau before it reaches the maximum theoretical values.

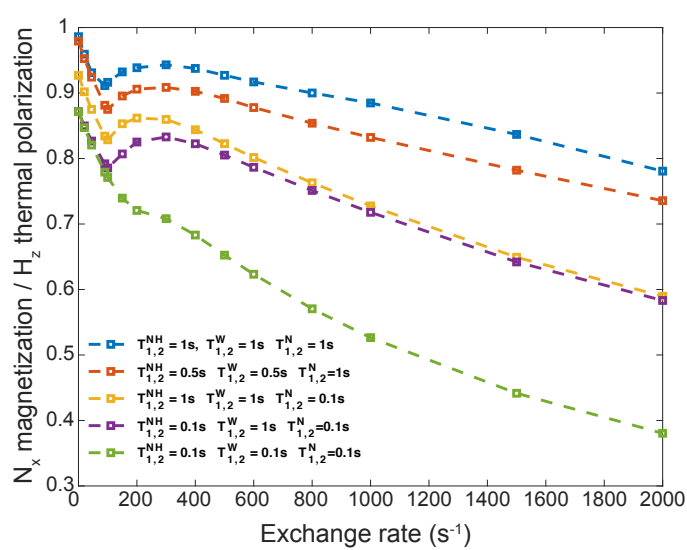

**Figure S6.** Maximum $N_X$ achieved within 100 ms of CP transfer, as a function of the $k_{HW}$ exchange rate. Different curves represent different sets of $^1H$ and $^{15}N$ relaxation parameters for protons and nitrogen. Notice the importance of water's relaxation: at fast exchange rates the labile proton acts as a mediator of polarization transfer, but the overall transfer efficiency is not dictated by their $T_1/T_2$, but rather by water's relaxation.

Another aspect worth discussing is how these features will change upon considering $^1H_n \rightarrow ^{15}N$ systems, where multiple labile $^1H$s can transfer polarization to the $^{15}N$. These aspects were examined with the aid of Liouville-space numerical simulations, to clarify the nitrogen buildups expected for $H_2N$s and $H_3N$s of the kind present in amines and protonated amines. Figure S7 compares these buildups for three relevant systems, in the absence and in the presence of fast chemical exchange. As known from the literature,[1] a multi-frequency interferogram arises in these multi-spin systems in the absence of exchange. As exchange sets in, however, the J-CP transfer dynamics of $H_2N$ and $H_3N$ systems become remarkably similar to that observed for HNs. In fact, for the $H_2N$ and $H_3N$ spin systems, the presence of additional polarization sources appears to improve the transfer efficiency even further than in spin-pairs. This is observable in the case of fast chemical exchange rate of 500 s$^{-1}$, when the maximum amplitude of $N_x$ reached much higher values at shorter contact times than in the case of HN spin system. This can be better appreciated in Figures S7b and S7c, which show the dependence of maximum $N_x$ amplitude and optimal contact time for which that buildup was reached.

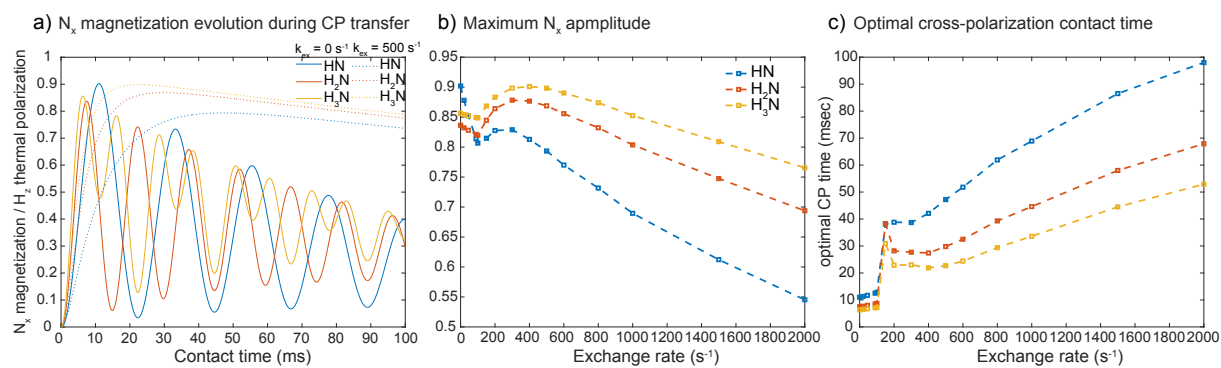

**Figure S7.** a) Amplitude of nitrogen buildup with three different spin systems in the absence and in the presence of fast chemical exchange. b) Maximum $N_x$ amplitudes achievable for different chemical exchange rates and c) corresponding contact times needed for these amplitudes to be reached.

**Supporting Information 4: On the accuracy of the Spinach exchange model**

An item to decide in the Liouville-space simulations, concerned how to account for the water pool. This was eventually considered as an ensemble of many independent spins; this raised the issue of which number of spins were needed to simulate an abundant water pool that can mimic a realistic repolarization, but remaining within the confine of the computationally practical. Varying the number of water protons in the system and measuring the transfer efficiency was one way of evaluating this; we observed that 50 water protons were sufficient to reproduce the experimental data, with the results of the simulations plateauing after exceeding ca. 150 proton systems (Figure S8). All our calculations thus used 200 water protons, to be reliable without demanding too much computing power: all simulations were performed on regular laptops, within reasonable times. While with 50 water protons, 100 ms CP simulation can be done in just a minute, 200 water protons increase the processing time to tens of minutes. Figure S8 further analyses the case of J-CP for long contact times, showing they have similar effects as looping short contacts multiple times as discussed in the main text.

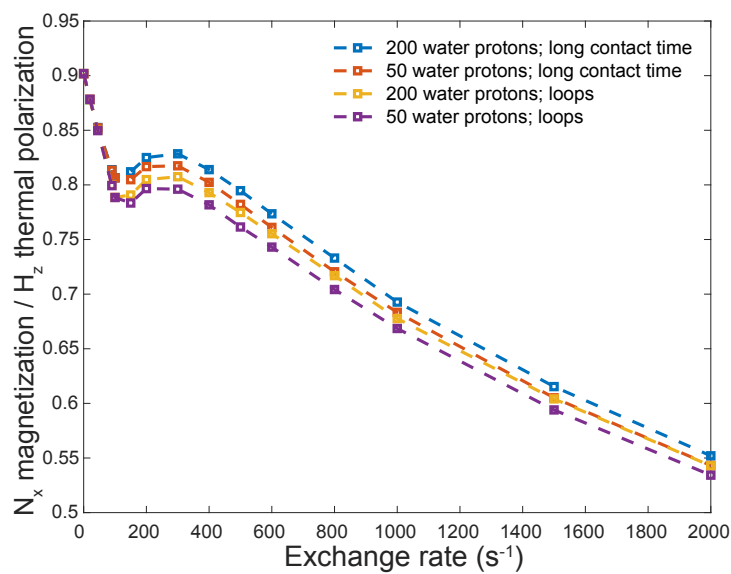

**Figure S8.** Comparison between 50 and 200 water protons in the system confirms that there is not a significant difference between the buildup efficiencies. Besides single-contact CP with optimal long times, these calculations involved looping multiple J-CPs with 11.1 ms per contact ($1/J_{NH}$). The 1:1 correlation between these two sets of results confirms that 200 water protons suffice, and that if water is effectively spin locked, it constantly repolarizes labile protons without the need to store and re-excite the protons multiple times in shorter J-CP loops.

## Supporting Information 5: On the use of L-CCP to polarize multiple reservoirs simultaneously

The possibility of utilizing the L-CCP procedure to polarize multiple $^{13}C$ simultaneously is briefly analyzed with the Liouville-space simulations shown in Figure S9, where the $^{15}N$ is assumed coupled to two carbons (akin to protein backbone coupling to CO and Ca). The $^{15}N \rightarrow ^{13}C$ portion of the transfer could involve a dual-band spin-lock on the carbon channel, or an alternate procedure where one shifts the carbon offset of a conventional CP in every consecutive loop back and forth between CO and Ca. Either scheme would provide an efficient buildup of both carbon magnetizations; in the present case simulations simply assumed that $\omega_1$ was much bigger than either offset.

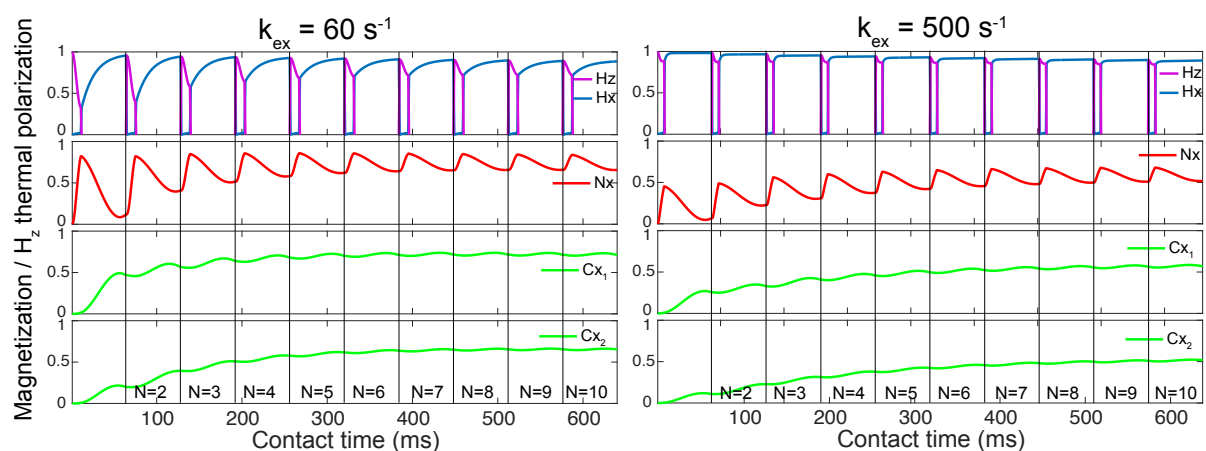

**Figure S9.** Numerical simulations showing the magnetizations of a four-spin system ($H^N$, $^{15}N$ and $^{13}C_1$, $^{13}C_2$), immersed in a bath of 200 additional water $^1Hs$ chemical exchanging with $H^N$, throughout multiple loops of the pulse sequence. Magnetization trajectories only show x components (except for protons where interleaved x and z components are shown), for $k_{HW}$ exchange rates of 60 s$^{-1}$ and 500 s$^{-1}$. $J_{NH}$ was set to 90 Hz, while $J_{NC}$ were 16 Hz and 11 Hz for the carbonyl and alpha carbon respectively. The N=1 loop would correspond to a conventional CCP

## References

(1) Bertrand, R. D.; Moniz, W. B.; Garroway, A. N.; Chingas, G. C. Carbon-13-Proton Cross-Polarization in Liquids. *J. Am. Chem. Soc.* **1978**, *100* (16), 5227–5229. https://doi.org/10.1021/ja00484a063.